\documentclass[journal]{IEEEtran}
\usepackage{amsthm}
\usepackage{amsfonts}
\usepackage{amssymb}
\ifCLASSINFOpdf
  \usepackage[pdftex]{graphicx}
  \usepackage{graphicx}
  \graphicspath{{../pdf/}{../jpeg/}{../png/}}
  \DeclareGraphicsExtensions{.pdf,.jpeg,.png}
\else
  \usepackage[dvips]{graphicx}
  \graphicspath{{../eps/}}
  \DeclareGraphicsExtensions{.eps}
\fi
\ifCLASSOPTIONcompsoc 
  \usepackage[caption=false,font=normalsize,labelfont=sf,textfont=sf]{subfig}
\else
\usepackage[caption=false,font=footnotesize]{subfig}
\fi
\usepackage{multirow}
\usepackage{epstopdf}
\usepackage{float}
\usepackage{balance}
\usepackage{xcolor}

\usepackage[hidelinks]{hyperref}
\usepackage{cite}
\ifCLASSINFOpdf
   \usepackage[pdftex]{graphicx}
   \else
\usepackage[dvips]{graphicx}
\fi
\usepackage[cmex10]{amsmath}
\usepackage[T1]{fontenc}
\newcommand*{\affmark}[1][*]{\textsuperscript{\dag}}
\usepackage{algorithm}
\usepackage[noend]{algpseudocode}
\makeatletter
\def\BState{\State\hskip-\ALG@thistlm}
\makeatother

\begin{document}
\title{Semi-Blind Channel Estimation and Hybrid Receiver Beamforming in the Tera-Hertz Multi-User Massive MIMO Uplink}
\author{
\normalsize{Abhisha~Garg~\IEEEmembership{Graduate Student Member,~IEEE,} Suraj~Srivastava,~\IEEEmembership{Member,~IEEE,} Varsha~Dubey, Aditya~K.~Jagannatham,~\IEEEmembership{Senior Member,~IEEE} and Lajos Hanzo,~\IEEEmembership{Life Fellow,~IEEE}\vspace{-1.8\baselineskip}}
\thanks{The work is supported by IEEE SPS scholarship grant for $2023, 2024$ and $2025$. The work of Aditya K. Jagannatham was supported in part by the Qualcomm Innovation Fellowship; in part by the Qualcomm 6G UR Gift; in part by the Arun Kumar Chair Professorship. The work of S. Srivastava was supported in part by Anusandhan National Research Foundation’s PM-ECRG/2024/478/ENS; and in part by Telecom Technology Development Fund (TTDF) under Grant TTDF/6G/368. S. Srivastava and A. K. Jagannatham jointly acknowledge the funding support provided by Anusandhan National Research Foundation's Advanced Research Grant ANRF/ARG/2025/005895/ENS. The work of Lajos Hanzo is supported by Engineering and Physical Sciences Research Council (EPSRC) projects is gratefully acknowledged: Platform for Driving Ultimate Connectivity (TITAN) (EP/X04047X/1; EP/Y037243/1); Robust and Reliable Quantum Computing (RoaRQ, EP/W032635/1); PerCom (EP/X012301/1); India-UK Intelligent Spectrum Innovation ICON UKRI-1859. S. Srivastava, A. K. Jagannatham and Lajos Hanzo jointly acknowledge the funding support provided to ICON-project by DST and UKRI-EPSRC under India-UK Joint opportunity in Telecommunications Research.

Abhisha Garg and Aditya K. Jagannatham are with the Department of Electrical Engineering, Indian Institute of Technology Kanpur, Kanpur-208016, India (e-mail: abhisha20@iitk.ac.in; adityaj@iitk.ac.in).

Suraj Srivastava is with the Department of Electrical Engineering, Indian Institute of Technology Jodhpur, Jodhpur, Rajasthan 342030, India (e-mail: surajsri@iitj.ac.in).

Varsha Dubey is with Apple Inc, Bengaluru, 560001, India. 

L. Hanzo is with the School of Electronics and Computer Science, University of Southampton, Southampton SO17 1BJ, U.K. (email:lh@ecs.soton.ac.uk)}}
\maketitle
\begin{abstract}
We develop a pragmatic multi-user (MU) massive multiple-input multiple-output (MIMO) channel model tailored to the THz band, encompassing factors such as molecular absorption, reflection losses and multipath diffused ray components. Next, we propose a novel semi-blind based channel state information (CSI) acquisition technique i.e. MU whitening decorrelation semi-blind (MU-WD-SB) that exploits the second order statistics corresponding to the unknown data symbols along with pilot vectors. A constrained Cramér-Rao Lower Bound (C-CRLB) is derived to bound the normalized mean square error (NMSE) performance of the proposed semi-blind learning technique. Our proposed scheme efficiently reduces the training overheads while enhancing the overall accuracy of the channel learning process. Furthermore, a novel hybrid receiver combiner framework is devised for MU THz massive MIMO systems, leveraging multiple measurement vector based sparse Bayesian learning (MMV-SBL) that relies on the estimated CSI acquired through our proposed semi-blind technique relying on low resolution analog-to-digital converters (ADCs). Finally, we propose an optimal hybrid combiner based on MMV-SBL, which directly reduces the MU interference. Extensive simulations are conducted to evaluate the performance gain of the proposed MU-WD-SB scheme over conventional training-based and other semi-blind learning techniques for a practical THz channel obtained from the high-resolution transmission (HITRAN) database. The metrics considered for quantifying the improvements include the NMSE, bit error rate (BER) and spectral-efficiency (SE).
\end{abstract}
\IEEEpeerreviewmaketitle
\section{Introduction}
Tera-Hertz (THz) schemes might find employment in next-generation wireless systems due to their potential for ultra-high data rates \cite{akyildiz2022terahertz}. The abundance of hither to unused spectrum available in the THz band, spanning from ($0.1-10$) THz \cite{jornet2011channel}, might meet the growing data-rate requirements. However, the practical deployment of THz communication faces considerable challenges attributed to substantial atmospheric turbulence and path losses \cite{piesiewicz2007scattering}. These challenges arise primarily because the high carrier frequency suffers from gas absorption effects. A promising technique of facing these challenges involves multiple-input multiple-output (MIMO) solutions relying on antenna arrays \cite{yue2023hybrid}. These arrays have the potential to enhance signal reception by forming highly directional beams with exceptionally high gains \cite{wu2023simultaneous} for mitigating the losses occuring in the THz band. At this juncture, it is also important to highlight that the traditional transceiver design, which requires a dedicated radio frequency (RF) chain for each antenna, tends to result in a significant escalation of hardware complexity. Therefore, the hybrid transceiver architecture, introduced by Molish \textit{et al.}, in their groundbreaking research \cite{molisch2017hybrid}, emerges as an appealing solution for THz systems, enabling the implementation of a moderate-complexity transceiver using a limited number of RF chains. Additionally, a hybrid MIMO transceiver strategically partitions the signal processing tasks between the analog RF front-end and the digital baseband processor \cite{srivastava2022hybrid}. Therefore, the holistic performance of the hybrid architecture hinges on the meticulous design of the baseband and RF transmit precoders/ and receiver combiners, which in turn, critically depend on the CSI accuracy \cite{xing2023unified}. Thus, achieving high-accuracy channel estimation is critical for constructing robust THz massive MIMO systems. Furthermore, a notable drawback of training-based channel estimation techniques is the reduction in spectral efficiency. By contrast, blind estimation methods eliminate the pilot overhead, but suffer from increased computational complexity and convergence-related problems. Hence, semi-blind techniques emerge as a promising alternative to both blind and training-based methods \cite{srinivas2019iterative}. They are capable of improving the channel estimation accuracy by leveraging the statistical characteristics of the THz system, while minimizing the pilot overhead. In the following section, a concise overview of the relevant prior research is provided.
\vspace{-3mm}
\subsection{Review of existing works}
High-frequency THz bands result in substantial propagation losses, including molecular absorption losses in terrestrial communication, necessitating the use of a significant number of small antenna elements to achieve high beamforming gains. Jornet and Akyildiz \cite{jornet2011channel} in their cutting-edge work, developed a THz channel model that incorporates propagation and molecular absorption losses. This model has proven to be beneficial for quantifying the capacity of THz systems. In their pioneering work, Piesiewicz \textit{et al.} \cite{piesiewicz2007scattering} considered indoor THz scenarios, specifically focusing on the calculation of reflection losses in the specular direction within non-line-of-sight (NLoS) scenarios. They conclusively demonstrated that scattering has a substantial impact on both the propagation characteristics and on absolute power levels in these scenarios. Transmit power variation with respect to communication distance has been investigated in \cite{zhang2025tensor},\cite{zhang2024channel} offering key insights into how distance-dependent path loss and beamforming influence the power requirements in sub-THz/THz systems. In their work, Dovelos \textit{et al.} \cite{dovelos2021channel} relied solely on the utilization of training symbols for THz channel estimation. Wei \textit{et al.} \cite{wei2022accurate}, proposed a pilot-based three-stage channel estimation scheme for a THz massive MIMO system. In stage-$1$, the orthogonal matching pursuit (OMP) algorithm is utilized for channel estimation, followed by stage-$2$, where the AoA/AoD estimate is refined using the expectation maximization (EM) algorithm. Finally, in stage-$3$, the path delays are estimated using the dictionary constructed. However, training symbols do not convey any information and their transmission results in a reduction of effective spectral efficiency. Thus, Mao and Wang \cite{mao2021terahertz} proposed a novel pilot design that uses index modulation (IM). This approach allows for the dynamic adjustment of pilot positions within the data frame and enables the transmission of additional information bits through the indices associated with these pilots, showing significant performance improvement over pilot based channel estimation. The early work of Abuthinien \textit{et al.} \cite{abuthinien2008semi}, incorporated the data symbols into the estimation process, which significantly improved the estimation accuracy and reduces the overall training overhead. To leverage the advantages of the data symbols, Liu \textit{et al.} \cite{liu2022blind} proposed subspace-based blind estimation in a Poisson distributed scenario, relying on second-order statistics. However, it is worth noting that blind techniques are frequently associated with excessive computational complexity and convergence related issues. Hence, to bridge the gap between pilot-based and blind estimation techniques, this work motivates the design of novel semi-blind techniques for MU THz massive MIMO systems by leveraging both the pilot and data symbols for enhancing the channel estimation performance.

Moreover, in an MU THz system with single-antenna nodes, purely blind estimation often suffers from a lack of diversity due to severe path loss, fewer multipath components, and narrow beam coverage at THz frequencies. With multiple users sharing the spectrum, these constraints can lead to ambiguous or unstable channel estimates when relying solely on statistical features \cite{aldana2003channel}. This lack of diversity can cause slow convergence or outright estimation failures \cite{nayebi2017semi}. The extremely high bandwidth also imposes stringent requirements on the estimation accuracy and makes purely blind methods prone to errors. As a result, semi-blind or pilot-based schemes become more attractive by providing partial reference data to anchor estimates, thus ensuring more robust and reliable operation in THz scenarios. However, semi-blind estimation offers a balance between blind and pilot-based estimation techniques by leveraging both pilot symbols and received data for improved channel estimation. Compared to blind estimation, semi-blind techniques reduce computational complexity by avoiding the need for fully unsupervised optimization, which often requires iterative algorithms such as EM \cite{nayebi2017semi} or high-dimensional subspace decomposition \cite{li2003subspace}. Conversely, compared to pilot-based methods, semi-blind estimation reduces pilot overhead by utilizing the statistical structure of the received data to refine channel estimates, which is particularly beneficial for high-dimensional systems such as THz massive MIMO \cite{muquet2002subspace}.

The versatility and relevance of semi-blind techniques is not limited to conventional MIMO orthogonal frequency division multiplexing (MIMO-OFDM) technologies, it can be effectively applied also in Filter Bank Multicarrier (FBMC) systems. The statistical properties of the signal and interference components within an OFDM-OQAM (Offset Quadrature Amplitude Modulation) system were harnessed by Su \cite{su2015semiblind} through the use of semi-blind techniques. His work underscored the efficacy of these techniques, particularly when dealing with scenarios characterized by a limited coherence time. Furthermore, Hou and Champagne \cite{hou2014semiblind} addressed the challenges posed by time-dispersive fading channels for OFDM-OQAM systems. The authors applied low-rank filtering to the channel coefficients to mitigate semiblind estimation noise. The authors in \cite{wan2011semiblind} proposed a novel semi-blind sparse channel learning technique that leverages the second-order statistics of the received signal in MIMO-OFDM systems.

The treatise \cite{chen2009semiblind} proposed a ground-breaking semi-blind learning technique for a single-carrier (SC) massive MIMO system, followed by the optimal design for the precoding sequence. Chu \textit{et al.} \cite{chu2018semi} contributed to the semi-blind estimation literature by conceiving convex optimization algorithms. Specifically, their estimator combined the atomic norm minimization (ANM) and $l_1$-minimization techniques to improve the performance of the mmWave MIMO system attained. The semi-blind algorithms may also be further extended by harnessing an EM framework, as demonstrated by the pioneering work of Nayebi and Rao \cite{nayebi2017semi}. Explicitly, they developed a pair of EM-based algorithms for semi-blind estimation in Time Division Duplex (TDD) MU MIMO systems. Their work clearly demonstrated the benefits of the SB channel estimation method for both uplink and downlink transmission scenarios. Abdallah and Darya \cite{abdallah2020semi} applied the EM framework in conjunction with a pair of decision-directed estimation strategies for semi-blind estimation in the specific context of molecular communication. Furthermore, Al-Shoukairi and Rao \cite{al2021semi} introduced an eigenvalue decomposition (EVD) based technique for reducing the dimensionality of the EM algorithm and addressed the semi-blind estimation problem of MIMO systems. 

In their cutting-edge work, Zhang \textit{et al.} \cite{zhang2014two} introduced a novel near-capacity MIMO system with norm-based joint transmit and receive antenna selection. The approach features a semi-blind process, combining a low-complexity minimum mean square error (MMSE)-based channel estimator with turbo detection. This enables the system to approach the performance of the maximum-likelihood (ML) estimator despite its significantly lower complexity. Xing \textit{et al.}, \cite{xing2023unified} proposed a joint matrix-monotonic optimization framework, for optimizing the linear transmit precoders and training sequences in MIMO systems considering both statistical and estimated CSI. Zhang \textit{et al.}, \cite{zhang2013reduced} in their ground-breaking work proposed a low-complexity joint CSI and three-stage iterative demapping scheme for MIMO systems. Their semi-blind approach utilized minimal training blocks to achieve near-optimal ML turbo detection performance with low computational complexity. An iterative semi-blind channel estimation technique for pilot contaminated massive MIMO systems has been advanced in \cite{srinivas2019iterative}. Rekik \textit{et al.} \cite{rekik2024fast} proposed a subspace-based semi-blind channel estimation method, where the covariance matrix and noise subspace are estimated for each subcarrier by exploiting the orthogonality of OFDM systems. The channel coefficients are subsequently estimated by minimizing the global cost function. The authors of \cite{karataev2024bilinear} proposed a semi-blind joint channel estimation and data detection framework for cell-free massive MIMO systems. Although the literature on semi-blind estimation extensively covers conventional terrestrial systems and scenarios, to the best of our knowledge, this is the first paper exploring the advantages of semi-blind estimation techniques in the THz domain.
\begin{table*}
    \centering
   
\caption{\small Contrasting the salient contributions of the present work to the literature} \label{tab:lit_rev}

\scriptsize
\begin{tabular}{|l|c|c|c|c|c|c|c|c|c|c|c|c|c|}

    \hline

\textbf{Features}  & \cite{dovelos2021channel} &\cite{nayebi2017semi} &\cite{abdallah2020semi}& \cite{ding2018bayesian}  &\cite{ma2022data} & {\cite{wei2022accurate}} & {\cite{rekik2024fast}} & {\cite{karataev2024bilinear}} & {\cite{srinivas2019iterative}} & {\cite{loukil2020terahertz}} & {\cite{zhang2021analysis}} & {\cite{zhao2024dynamic}} &\textbf{This paper} \\

 \hline

THz Band

& \checkmark   &  &  &  &  & \checkmark & & & & \checkmark & \checkmark & \checkmark & \checkmark\\

 \hline
 
MU massive MIMO

 &   & \checkmark &  & \checkmark & \checkmark &  &  & & \checkmark & & \checkmark & \checkmark & \checkmark \\

 \hline

 {Semi-blind }

 &   &   &  &   & &  & \checkmark & \checkmark & \checkmark & & & & \checkmark\\

\hline

 {Pilot-based}

 &   &   &  &   & & \checkmark &  &  &  & & & & \\

 \hline

Reflection \& Molecular Absorption losses
   & \checkmark   &   &  &   & & &  &  &  & & \checkmark & & \checkmark\\

 \hline

Constrained CRLB

&  & \checkmark & \checkmark &   & & &  &  &  & & & & \checkmark\\
 \hline

 Low Resolution ADCs in THz band

  &  &  &  &  & & &  & &  & & \checkmark & \checkmark & \checkmark\\

 \hline

\textbf{Diffused-ray modeling}

   &     &   &  &  & &  &  &  &  & & & & \checkmark\\
 \hline

\textbf{Semi-blind CSI acquisition in THz band}
  &  & & &  & &  &  &   &  & & & & \checkmark\\

 \hline

\textbf{Hybrid receiver combiner using semi-blind estimated CSI}

    &      &   &  &  & & & & &  & & & & \checkmark\\
 \hline

\end{tabular}
\end{table*}

Recent advances in THz massive MIMO-ISAC systems leverage tensor-based techniques for joint channel estimation and target sensing \cite{zhang2024channel, zhang2024integrated}. These works employ structured tensor decomposition methods, such as canonical polyadic decomposition (CPD), to estimate multi-path parameters in the angular, delay, and Doppler domains, while enabling efficient channel reconstruction with minimal pilot overhead. Additionally, segment-based training patterns have been introduced to mitigate beam squint effects, improving sensing resolution and estimation accuracy. While tensor decomposition-based methods efficiently exploit high-dimensional channel structures, they impose strict sparsity and rank constraints and require high computational resources. By contrast, the proposed semi-blind technique combines pilot and data symbols, reducing reliance on such structural assumptions while maintaining computational efficiency, making it a more adaptable solution for THz channel estimation in practical scenarios subject to a constrained pilot overhead. Moreover, the recent study \cite{zhang2025terahertz} provides a comprehensive survey of THz–ISAC empowered unmanned aerial vehicles (UAVs), analyzing propagation characteristics and system-level challenges, with a focus on hybrid beamforming, waveform design, and CSI acquisition techniques in such systems.

In practical scenarios, the importance of channel estimation is evident in the design of hybrid beamformers, which motivates us to develop a novel hybrid combiner framework for analog/digital beamforming. Additionally, THz systems tend to rely on substantial number of antenna elements, hence leading to a large channel matrix. To address the associated hardware-related challenges, low-resolution ADCs play a crucial role in designing the hybrid transceivers for real-world applications \cite{ding2018bayesian}. The pioneering work of Ayach \textit{et al.} \cite{el2014spatially} leveraged the spatial characteristics of mmWave channels to frame the hybrid transceiver problem as a sparse reconstruction problem and provide an algorithmic solution using the popular orthogonal matching pursuit (OMP). Alkhateeb \textit{et al.} \cite{alkhateeb2013hybrid} tackle the problem of hybrid beamforming in single user (SU) mmWave systems, where they rely on partial channel knowledge at the BS and MS, exploiting solely the angle of arrival (AoA) information.  Morsali and Champagne \cite{morsali2019robust} proposed a robust hybrid transceiver design for the uplink of massive MIMO systems, while considering realistic imperfect CSI. Ma \textit{et al.} \cite{ma2022data} employed a hybrid beamforming in mmWave MU-MIMO systems in support of user diversity. Their approach combined the array and spatial signal processing techniques to form highly directional beams having substantial beamforming gains. Furthermore, none of the existing solutions considers the THz design of the hybrid combiner using the estimated channel harnessing low-resolution ADCs. To fill this knowledge gap, the next section provides a brief overview of the main contributions of our work. Furthermore, Table-\ref{tab:lit_rev} boldly contrasts our contributions to the existing literature.
\subsection{Novel contributions of the paper}
{\begin{enumerate}
\item An end-to-end THz channel model has been developed considering low-resolution ADCs and diffused ray modeling. Motivated by the challenges in accurate channel estimation under these conditions, this work introduces a semi-blind Regularized Alternating Least Squares (RALS)-based channel estimation framework. The RALS-SB channel estimator leverages both pilot and data symbols to achieve enhanced performance.
\item To further improve the NMSE performance and eliminate dependence on the regularization parameter, this paper introduces a novel semi-blind approach termed the MU whitening decorrelation semi-blind (MU-WD-SB) technique. This technique involves a two-step process: the first step is blind estimation, which solely leverages the second-order statistics of the transmitted data symbols, followed by channel estimation using pilot symbols.
\item The accuracy of our proposed semi-blind estimator is characterized by comparing the estimation error variance with the C-CRLB. While a training-based method requires a substantial number of pilot symbols for enhancing the estimation accuracy, our proposed semi-blind scheme exploits the data symbols, hence reducing the pilot overhead required for achieving an equivalent estimation performance. In essence, for the same pilot length, our SB approach provides a superior accuracy compared to training-based techniques.
\item Furthermore, we have designed a hybrid receiver combiner (RC) based on the estimated CSI to evaluate the capacity in a system employing low-resolution ADCs. Our simulation results illustrate the improved performance of the proposed semi-blind estimators and hybrid RC, across a range of practical simulation parameters. Furthermore, the various channel impairments encountered in the THz band are characterized by relying on the HIgh resolution TRANsmission (HITRAN) database \cite{hill}, which renders the model practically viable.
\end{enumerate}}
\subsection{Organization of the paper}
In Section-II we commence by portraying over MU massive MIMO THz system and channel model, taking into account the critical aspects of a typical THz channel. Subsequently, in Section-III we highlight the conventional channel estimation scheme MU THz ML i.e the MU-THz-ML and the RALS based semi-blind framework i.e the MU-RALS-SB algorithm. This is followed in Section-IV by proposing our novel semi-blind framework, i.e the MU-WD-SB algorithm, that leverages both the data and pilot symbols. Section-V proceeds with the derivation of the C-CRLB for the proposed MU-WD-SB technique, along with an assessment of the MSE improvement observed over the conventional ML based technique when employing our MU-WD-SB technique. In Section-VI, we introduce the hybrid combiner based on the MMV-SBL technique advocated. Section-VII includes extensive simulation results, while Section-\ref{conclusion} provides our conclusions along with supplementary evidence supporting several findings in the Appendices.
\subsection{\textbf {Notation:}} Matrices and vectors are denoted by the capital and lower case bold face letters $\mathbf{A}$ and $\mathbf{a}$, respectively. The superscripts $(\cdot)^H$, $(\cdot)^T$,$(\cdot)^*$, $(\cdot)^{-1}$ and $(\cdot)^{\dagger}$ represent the  Hermitian, conjugation, transpose, inverse and  pseudo-inverse of a matrix, respectively. The statistical expectation operator is denoted by $\mathbb{E}\{{\cdot}\}$ and symbol $\mathbb {B}$ denotes a set of binary values. The symbols $\otimes$ and $\mathrm{Tr}(\cdot)$, respectively, stand for the matrix Kronecker product and the trace operator. The Frobenius norm of a matrix is represented by  $\|\cdot\|_{\mathcal{F}}$, $\|\cdot\|_0$ denotes the $l_0$ norm, while $\text{vec}(.)$ represents the vectorize operator, where $\text{vec}(\mathbf{ABC}) = (\mathbf{C}^T \otimes \mathbf{A}) \text{vec}(\mathbf{B})$.
\section{MU Massive MIMO THz System and Channel Models}
\begin{figure}
\centering
\includegraphics[scale=0.25]{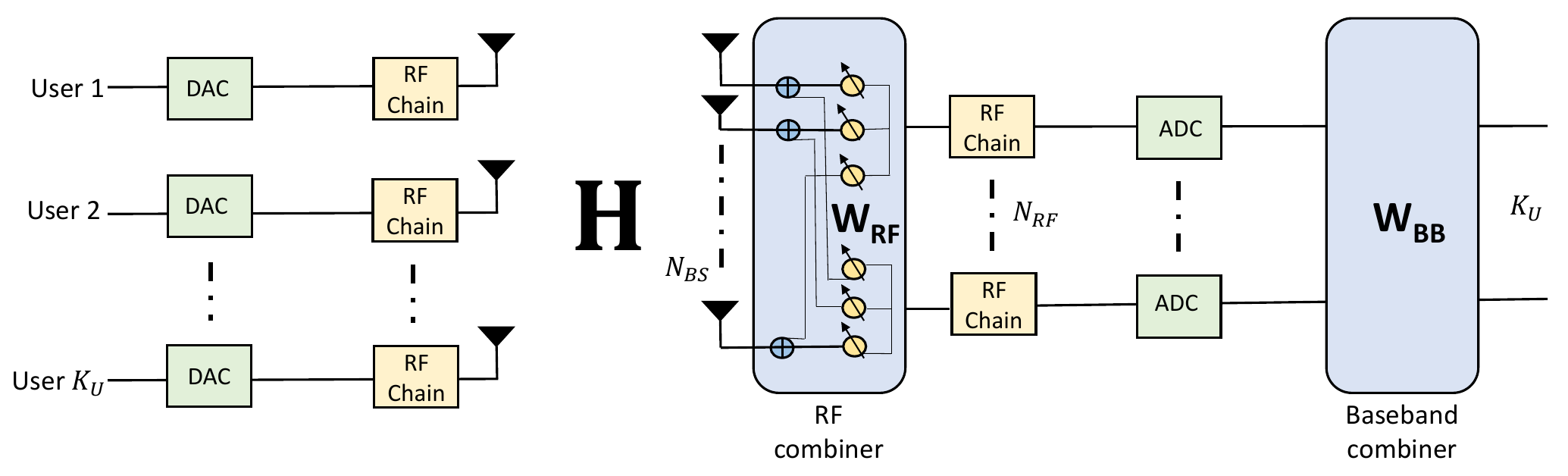}
\vspace{-24pt}
\caption{Block diagram of a MU THz massive MIMO system.}
\label{fig:schematic-diag}
\end{figure}

We consider a single THz BS which is equipped with a uniform linear array (ULA) containing $N_{BS}$ receiver antenna (RA) elements serving $K_U$ spatially distributed uplink (UL) transmitter each having a single transmitter antenna (TA). The BS employs a hybrid MIMO architecture equipped with $N_{RF}$ RF chains, which follows the relationship of $K_U \leq N_{RF} \ll N_{BS}$, as seen in Fig. \ref{fig:schematic-diag}.
The received signal at each RA is converted to the digital domain by a network of phase shifters in a fully connected fashion, which is represented by the analog RF RC matrix $\mathbf {W}_{\text {RF}} \in \mathbb {C}^{N_{BS}  \times  N_{RF}}$. Importantly, the constant modulus constraint of $|\mathbf{W}_{\text {RF}}(\rho,\kappa)| = 1/\sqrt{N_{BS}}, 1 \leq \rho \leq N_{BS}, 1 \leq \kappa \leq N_{RF}$, is satisfied by the phase shifters. Subsequently, the conversion of the signal into $K_U$ distinct user streams is carried out by the digital baseband RC matrix denoted as $\mathbf {W}_{\text {BB}}\in \mathbb {C}^{N_{RF}  \times K_{U} }$.
\subsection{MU massive MIMO THz channel model}
The THz channel can be characterized as a composite of line-of-sight (LoS) and NLoS path components. Moreover, the LoS path signifies the direct link connecting the BS and the UE, while the NLoS paths encompass numerous multipath reflections due to the scatterers present in the environment. Thus, the composition of the channel spanning from the $k_u$th user to the BS comprises $(N_{\text{LoS}} + N_{\text{NLoS}})$ paths, where the relationship of $(N_{\text{LoS}} + N_{\text{NLoS}} \ll N_{BS})$ holds. Furthermore, the characterization of the $z$th path involves the complex path gain $\alpha_{k_u,z}$ and the corresponding angle of arrival (AoA) $\phi_{k_u,z}$. Thus, the array response vector associated with the ULA denoted as $\mathbf{a}(\phi_{k_u,z}) \in \mathbb {C}^{N_{BS}\times 1}$ is given by
\begin{multline}
    \mathbf{a}(\phi_{k_u,z}) = \frac{1}{\sqrt{N_{BS}}} \big[ 1, e^{-j\frac{2\pi}{\lambda}d_rcos({\phi_{k_u,z}})}, \cdots, \\ e^{-j\frac{2\pi}{\lambda}{(N_{BS}-1)}d_rcos({\phi_{k_u,z}})} \big]^T, \label{array-resp}
\end{multline}
where $\lambda$ denotes the operating wavelength and $d_r$ signifies the RA spacing. Therefore, the uplink THz channel is then formulated as
\begin{align}
\mathbf{h}_{k_u}(f,d) = \mathbf{h}_{\textrm{LoS},k_u}(f,d) + \mathbf{h}_{\textrm{NLoS},k_u}(f,d) \phantom{\beta} \forall \phantom{\beta} 1 \leq k_u \leq K_U, \label{channel-vec}    
\end{align}
where $f$ represents the operational frequency, while $d$ is the distance traversed by the ray. Additionally, one can articulate the LoS and NLoS constituents as \cite{garg2024angularly}
\begin{equation}
\mathbf{h}_{\textrm{LoS},k_u}(f,d) = \sqrt{N_{BS}}\alpha_{k_u}(f,d)B_r\mathbf{a}_r({\phi_{k_u}}), \label{LOS-channel}
\end{equation}
\begin{align}
\mathbf{h}_{\textrm{NLoS},k_u}(f,d_{z,\ell}) = \sqrt{\frac{N_{BS}}{N_{\textrm{NLoS}}N_{\textrm{ray}}}}\sum_{z=1}^{N_{\textrm{NLoS}}}\sum_{\ell=1}^{N_{\textrm{ray}}}\notag \\ \alpha_{k_u,z,\ell}(f,d_{z,\ell})B_r\mathbf{a}({\phi_{k_u,z,\ell}}).\label{NLOS-channel}
\end{align}    
Here, $N_{\textrm{ray}}$ represents the number of scattered rays within each multipath segment and $N_{\textrm{NLoS}}$ is the number of multipath constituents in the NLoS scenerio. The symbols $\alpha_{k_u}(f,d)$ and $\alpha_{k_u,z,\ell}(f,d_{z,\ell})$ denote the complex path gains for the LoS and NLoS components, respectively. Additionally, $\phi_{k_u}$ denotes the AoA within the LoS multipath segment, while $\phi_{k_u,z,\ell}$ is the AoA of the $\ell$th scattered ray within the $z$th NLoS multipath segment. Furthermore, $B_r$ represents the aggregate transmit and received antenna gain. The magnitude of the complex path gain $\alpha_{k_u}(f,d)$ for the LoS component can be further expressed as
\begin{align}
|\alpha_{k_u}(f,d)|^{2} = \mathit{L}_\textrm{spread}(\mathit{f},\mathit{d})\mathit{L}_\textrm{abs}(\mathit{f},\mathit{d}),
\label{los_path}
\end{align}
where $\mathit{L}_{\textrm{abs}}(\mathit{f},\mathit{d})$ represents the molecular absorption loss, as discussed in \cite{dovelos2021channel}, while $\mathit{L}_{\textrm{spread}}(\mathit{f},\mathit{d})$ represents the free-space loss. These parameters can be further characterized as
\begin{align}
\mathit{L}_{\textrm{spread}}(\mathit{f},\mathit{d}) = \left (\frac{\mathit{c}}{4\pi\mathit{f}\mathit{d}}\right)^{2}, \mathit{L}_{\textrm{abs}}(\mathit{f},\mathit{d}) = e^{-\mathit{k}_{\textrm{abs}}(\mathit{f})\mathit{d}}, \label{fre-space}
\end{align}
where $\mathit{k}_{\textrm{abs}}(\mathit{f})$ signifies the molecular absorption coefficient, and its specific definition is given by
\begin{align}
\mathit{k}_{\textrm{abs}}(\mathit{f}) = \sum_{\mathit{a},\mathit{g}}\mathit{k}_{\textrm{abs}}^{\mathit{a},\mathit{g}}(\mathit{f}). \label{abs_coeff}
\end{align}
The parameter $\mathit{k}_{\textrm{abs}}^{\mathit{a},\mathit{g}}(\mathit{f})$ represents the absorption coefficient pertaining to the $\mathit{a}$th isotopologue of the $\mathit{g}$th gas. It is worth noting that $\mathit{k}_{\textrm{abs}}^{\mathit{a},\mathit{g}}(\mathit{f})$ can be derived through calculations using the HITRAN database, as demonstrated in \cite{jornet2011channel}. Similarly, the mathematical formulation for the magnitude of the $\mathit{i}$th scattered ray, within the $\mathit{z}$th NLoS component, corresponding to the complex path gain of $\alpha_{k_u,z,\ell}(f,d_{z,\ell})$ can be formulated as
\begin{align}
|\alpha_{k_u,z,\ell}(f,d_{z,\ell})|^{2} = \Gamma_{z,\ell}^{2}(\mathit{f})\mathit{L}_{\textrm{spread}}(\mathit{f},\mathit{d}_{z,\ell})\mathit{L}_\textrm{abs}(\mathit{f},\mathit{d}_{z,\ell}), \label{Nlos_path}
\end{align}
where $\Gamma_{z,\ell}(f)$ denotes the first-order reflection coefficient  \cite{srivastava2022hybrid}. Furthermore, $\Gamma_{z,\ell}(\mathit{f})$ is defined as the product of the Fresnel reflection coefficient $\gamma_{z,\ell}(\mathit{f})$ and the Rayleigh roughness factor $\varrho_{z,\ell}(\mathit{f})$ \cite{piesiewicz2007scattering}. Their relationship can be expressed as
 \begin{align}
\Gamma_{z,\ell}(\mathit{f}) = \gamma_{z,\ell}(f)\varrho_{z,\ell}(\mathit{f}), \label{reflection_eqn}
\end{align}
\vspace{-15pt}
\begin{align}
\gamma_{z,\ell}(\mathit{f}) = \frac{Z(\mathit{f})\cos(\theta_{\text{in}_{z,\ell}}) - Z_{0}\cos(\theta_{{\text{ref}_{z,\ell}}})}{Z(\mathit{f})\cos(\theta_{{\text{in}_{z,\ell}}}) + Z_{0}\cos(\theta_{{\text{ref}_{z,\ell}}})},
\end{align}
\vspace{-12pt}
\begin{align}
\varrho(\mathit{f}) = e^{-\frac{1}{2}\left(\frac{4\pi\mathit{f}\sigma \cos\left(\theta_{{\text{in}_{z,\ell}}}\right)}{\mathit{c}}\right)^{2}}. \label{ray_rough}
\end{align}
The associated angle of refraction, denoted as $\theta_{{\text{ref}_{z,\ell}}}$ is defined by $\theta_{{\text{ref}_{z,\ell}}} = \sin^{-1}\left(\sin(\theta_{{\text{in}_{z,\ell}}})\frac{\mathit{Z}(\mathit{f})}{\mathit{Z}_{0}}\right)$, whereas $\theta_{{\text{in}_{z,\ell}}}$ represents the angle of incidence corresponding to $z$th path and $\ell$th diffuse ray. The wave impedance of free space is represented by \cite{piesiewicz2007scattering} $\mathit{Z}_{0} = 377\Omega$. Additionally, the wave impedance of the reflecting medium, $Z(\mathit{f})$ \cite{piesiewicz2007scattering}, is given by
\begin{align}
Z(\mathit{f}) = \sqrt{\frac{\mu_{0}}{\varepsilon_{0}\left(\mathit{n}^{2} - (\frac{\varsigma\mathit{c}}{4\pi\mathit{f}})^{2} - \mathit{j}\frac{2\mathit{n}\varsigma\mathit{c}}{4\pi\mathit{f}}\right)}}, \label{wav_imp}
\end{align}
where $\mu_{0}$ and $\varepsilon_{0}$ correspond to the constants representing the free-space permeability and permittivity, respectively. Additionally, $\varsigma$ denotes the absorption coefficient of the reflecting medium and $n$ signifies the index of refraction. The next subsection provides a concise overview of the received signal at the output of the RF RC.
\subsection{Pilot transmission}
Each user transmits a distinct orthogonal pilot signal, represented as $\mathbf{x}_{k_u} \in \mathbb{C}^{\tau_p \times 1}$ for the $k_u$-th user, where the pilot length $\tau_p$ satisfies the constraint $\tau_p \geq K_U$. The BS leverages these pilot signals to perform uplink channel estimation for all users. The pilot signal received at the BS undergoes processing through an analog RC matrix $\mathbf{W}_{\text{RF}}$, which leads to the signal $\mathbf{Y}_p \in \mathbb{C}^{N_{RF} \times \tau_p}$ in the digital front-end formulated as
\begin{equation} 
\mathbf{Y}_p=\mathbf {W}_{\text {RF}}^H \left(\sum_{k_u = 1}^{K_U} \mathbf{h}_{k_u}\mathbf{x}_{k_u}^H + \mathbf{V}_p \right), \label{system-model}
\end{equation}
\begin{figure}
\centering
\includegraphics[scale=0.9]{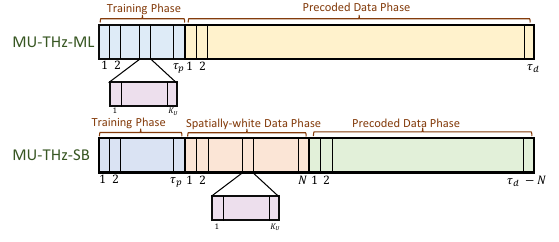}
\vspace{-14pt}
\caption{Frame structures of THz ML and SB channel estimation schemes}
\label{fig:frame-struc}
\end{figure}
where the components of the additive noise $\mathbf {V}_p$ are statistically independent and have an identical Gaussian distribution with zero mean and variance $\sigma_{v} ^{2}$. Because the pilot signals are orthonormal, we perform post-multiplication of $\mathbf{Y}_p$ with $\mathbf{x}_{k_u}$, in order to acquire the pilot signal required for estimating the channel of the $k_u$-th user, formulated as
\begin{equation}
\mathbf{y}_{k_u} = \mathbf{Y}_p\mathbf{x}_{k_u} = \mathbf{W}_{\text{RF}}^H\mathbf{h}_{k_u} + \mathbf{W}_{\text{RF}}^H\mathbf{V}_p\mathbf{x}_{k_u} \in \mathbb{C}^{N_{RF} \times 1}, \label{output_user}
\end{equation}
where the effective noise covariance of $\mathbf{v} = \mathbf{W}_{\text{RF}}^H\mathbf{V}_p\mathbf{x}_{k_u}$ is given as $\sigma_v^2\mathbf{W}_{\text{RF}}^H\mathbf{W}_{\text{RF}}$. It is apparent in \eqref{output_user} that the measurement vector $\mathbf{y}_{k_u}$ is devoid of interference from other users. It can be readily observed from \eqref{output_user} that the analog RC matrix $\mathbf{W}_{\text{RF}}$ plays a pivotal role in determining the channel estimation performance. Again, the analog RC matrix has to satisfy the constant modulus constraint, hence the elements of $\mathbf{W}_{\text{RF}}$ have to obey $ \frac{1}{\sqrt{N_{BS}}} \mathrm{exp}(j \psi)$, where the phase $\psi$ is selected randomly with a uniform probability distribution from the set $\mathcal{A} =\left\{\textrm{0},\frac{2\pi}{2^{N_{Q}}},\cdots,\frac{\left(2^{N_{Q}}-1\right)2\pi}{2^{N_{Q}}}\right\}$, with $N_Q$ being the quantization parameter. As described in \cite{prasanna2021mmwave}, this setting of $\mathbf{W}_{\text{RF}}$ during channel estimation ensures that $\mathbf{W}_{\text{RF}}\mathbf{W}_{\text{RF}}^H$ approaches $\mathbf{I}_{N_{BS}}$ as the number of antennas tends to infinity. The next subsection provides a succinct discussion of the conventional channel estimation technique.
\subsection{Conventional ML-based channel estimation}
In this subsection, we provide an overview of the conventional ML channel estimation technique, which forms our benchmarker. In order to estimate the MU THz channel, $\tau_p$ pilot beams are transmitted, as shown in Fig. \ref{fig:frame-struc} for training-based channel estimation. Upon concatenating the pilot vectors for all users including the complete set of pilot beams $\tau_p$, we obtain the transmit pilot matrix $ {\mathbf {X}_p} = \left [{ {\mathbf x}_{1}, {\mathbf x}_{2}, \ldots, {\mathbf x}_{K_U} }\right] \in\mathbb{C}^{\tau_p\times K_U}$. Again, the pilot vectors are orthogonal to one another, obeying
\begin{equation} \mathbf {X}_{p}^H{\mathbf{X}_p} = P_{p} \tau_p \mathbf {I}_{K_U},\label{pilot-orthog}\end{equation}
where $P_p$ denotes the pilot power. Therefore, the received pilot block before RF combining can be formulated as
\begin{align}
    \widetilde{\mathbf{Y}}_p = \mathbf{H}\mathbf{X}_p + \widetilde{\mathbf{V}}_p,
\end{align}
where $\mathbf{H} = [\mathbf{h}_1, \mathbf{h}_2, \cdots, \mathbf{h}_{k_u}] \in \mathbb{C}^{N_{BS} \times K_U}$ represents the MU THz channel matrix. To effectively excite different angular modes of the THz channel, we employ multiple combining matrices that capture the dominant spatial components. By leveraging different beamforming weight configurations, this approach ensures comprehensive exploration of the angular domain, which is crucial for accurate channel estimation and efficient beam alignment in highly directional THz communication systems. In this regard, let the combining matrix be divided such that $B = \frac{N_{BS}}{N_{RF}}$. Furthermore, let $\mathbf{W}_{\text{RF},b} \in \mathbb{C}^{N_{BS} \times N_{RF}}$ denote the RF combiner corresponding to the $b$-th sub-block, $1 \leq b \leq B$. Therefore, the received pilot matrix $\check{\mathbf{Y}}_{p,b} \in \mathbb{C}^{N_{RF} \times N_b}$ after applying the RF combiner can be formulated as
\begin{align}
    \check{\mathbf{Y}}_{p,b} = \mathbf{W}_{\text{RF},b}^H \mathbf{H} \mathbf{X}_p^H + \mathbf{W}_{RF,b}^H \widetilde{\mathbf{V}}_{p,b}.
\end{align}
Let the complete RF combiner matrix be defined as $\mathbf{W}_{\text{RF}} = [\mathbf{W}_{\text{RF},1}, \mathbf{W}_{\text{RF},2}, \cdots, \mathbf{W}_{\text{RF},B}] \in \mathbb{C}^{N_{BS} \times N_{BS}}$. Thus, the combined received signal across all the users for a pilot length of $\tau_p$ is given by
\begin{equation}
   \mathbf{Y}_p = {{\mathbf {W}}_{\text {RF}}^H}\mathbf {H}{\mathbf {X}_p^H} + \mathbf{W}_{\text{RF}}^H{\mathbf {V}}_p,\label{pilot-rx}
\end{equation}
The ML estimate derived for $\mathbf{H}$ is given as \cite{kay1993statistical}
\vspace{-5pt}
\begin{equation} \widehat {\mathbf {H}}_{\text {ML}} = \frac{1}{P_p \tau_p} \mathbf {W}_{\text {RF}}\mathbf {Y}_p (\mathbf{X}_p^H)^\dagger =\mathbf {H}+ \widetilde{\mathbf{V}}_p,\label{ML-est}\end{equation}
where we have $\widetilde{\mathbf{V}}_p = \frac{1}{P_p \tau_p} \mathbf{V}_p (\mathbf{X}_p^H)^\dagger$. It is worth noting that the ML estimate has a limitation; it does not exploit the information gleaned from the unknown data symbols. Hence, in order to improve the estimation accuracy of the ML scheme, one has to increase the number of pilots $\tau_p$, which in turn leads to increased pilot overheads and reduces the overall spectral efficiency. Thus, in the subsequent section, we introduce a RALS-SB based low-rank matrix completion approach for estimating the MU THz channel.
\begin{figure*}
\begin{align}
\widehat{\mathbf{V}}_{\text{A}}(:,i) & = \mathop{\text{argmin}}\limits_{{\mathbf{V}}_{\text{A}}(:,i)\in{\mathbb{C}}^{K_{U} \times 1}} \Vert {\mathbf{Y}}_{\omega,b}(:,i) - {\text{Diag}}[{\mathbf{W}}^H_{\text{RF}}(:,i)] {\widehat{\mathbf{U}}}_{\text{A}}\boldsymbol{\Lambda } {\mathbf {V}}_{\text{A}}(:,i)\Vert_{2}^{2} + \beta_{\text{V}} ~ \Vert{\mathbf{V}}_{\text{A}}(:,i) \Vert_{2}^{2} \notag \\ & = \left(\boldsymbol{\Lambda }{\widehat{\mathbf{U}}}_{\text{A}}^{H} ({\text{Diag}}[\mathbf{W}^H_{\text{RF}}(:,i)]){\widehat{\mathbf{U}}}_{\text{A}}\boldsymbol{\Lambda } + \beta_{\text{V}}{\mathbf{I}}\right)^{-1}\boldsymbol{\Lambda }{\widehat{\mathbf{U}}}_{\text{A}}^{H}({\text{Diag}}[{\mathbf{W}}_{\text{RF}}^H(:,i)]){\mathbf{Y}}_{\omega,b}(:,i). \label{expand_V} \end{align}
\vspace{-10pt}
\begin{equation}\widehat{\mathbf{U}}_{\text{A}}(n,:) = {\mathbf{Y}}_{\omega,b}(n,:)\left({\text{Diag}}[{\mathbf{W}}^H_{\text{RF}}(n,:)]\right){\widehat{\mathbf{V}}}_{\text{A}}^{H} \boldsymbol{\Lambda } \left(\boldsymbol{\Lambda }{\widehat{\mathbf{V}}}_{\text{A}}({\text{Diag}}[{\mathbf{W}}^H_{\text{RF}}(n,:)]){\widehat{\mathbf{V}}}_{\text{A}}^{H}\boldsymbol{\Lambda }+ \beta_{\text{U}}{\mathbf{I}}\right)^{-1}. \label{expand_U}\end{equation}
\hrulefill
\end{figure*}
\section{Semi-Blind Estimation using our RALS-SB approach}\label{RALS-derive}
In this section, we introduce a novel semi-blind approach that aims for substantially reducing the training overhead, relying on a low-rank matrix completion formulation. The low-rank property harnessed may be attributed to the observation that the number of users $K_U$ is significantly lower than the number of BS RAs $N_{BS}$, which is a typical scenario in MU THz systems. Let $\mathbf {X}_{d,b}\in \mathbb {C}^{N \times K_{U}}$ represent the transmit data matrix, comprising spatially uncorrelated data symbols. This matrix adheres to the condition $\mathbb {E}[{\mathbf{X}_{d,b}^H}\mathbf {X}_{d,b}]= P_{d} N \mathbf {I}_{K_U}$. Therefore, the corresponding received signal matrix $\mathbf{Y}_{d,b} \in \mathbb{C}^{N_{RF} \times N}$, at the output of the RF RC is given as
\begin{align}
    \mathbf{Y}_{d,b} = \mathbf{W}_{\text{RF}}^H \mathbf{H} {\mathbf{X}_{d,b}^H} + \mathbf{W}_{\text{RF}}^H \mathbf{V}_{d,b} \;, \label{data-model}
\end{align}
where the components of the additive noise $\mathbf{V}_{d,b} \in \mathbb{C}^{N_{BS} \times N}$ are independent and identically distributed (i.i.d) complex Gaussian with mean zero and variance $\sigma^2$. Moreover, in order to create the integrated pilot and data model, one can further concatenate Eq. \eqref{pilot-rx} and \eqref{data-model} to form $\mathbf{Y}_{\omega,b} = [\mathbf{Y}_p, \mathbf{Y}_{d,b}] \in \mathbb{C}^{N_{BS} \times (\tau_p+N)}$, which is given as 
\begin{align}
    \mathbf{Y}_{\omega,b} = \mathbf{W}_{\text{RF}}^H\mathbf{H}[\mathbf{X}_p^H, \mathbf{X}_{d,b}^H] + \mathbf{W}_{\text{RF}}^H[\mathbf{V}_p,\mathbf{V}_{d,b}].
\end{align}
Let the equivalent pilot-data matrix be $\mathbf{X}_b = [\mathbf{X}_p^H, \mathbf{X}_{d,b}^H]$, and the equivalent noise matrix be $\mathbf{V}_b = [\mathbf{V}_p,\mathbf{V}_{d,b}]$. Furthermore, for notational simplicity, let $\tau_c = \tau_p + N$. Consequently, the received concatenated model in compact form is given as
\begin{align}
    \mathbf{Y}_{\omega,b} = \mathbf{W}_{\text{RF}}^H\mathbf{H}\mathbf{X}_b + \mathbf{W}_{\text{RF}}^H \mathbf{V}_b.
\end{align}
Here, we introduce a semi-blind estimation approach based on RALS-SB for the recovery of $\mathbf{H}$ and $\mathbf{X}_{d,b}$ from $\mathbf{Y}_{\omega,b}$. Based on the above discussion, the problem of low-rank matrix completion is expressed as \cite{jain2013low}
\begin{align}\label{lowrank-prob} 
&\hspace {-.5pc}({\tilde{\mathbf {U}}_{\text{opt}}} , \tilde{{\mathbf {V}}}_{\text{opt}}) = \mathop { \mathop {\mathrm {argmin}}} \limits _{\substack{\scriptstyle {\mathbf {U}_{\text{A}}} \in \mathbb {C}^{N_{RF} \! \times \! K_U} \\ \scriptstyle {\mathbf {V}_{\text{A}}} \in \mathbb {C}^{K_U \! \times \! \tau_{c}} }} \left \|{ {\mathbf {Y}}_{\omega,b } - {\mathbf {U}_{\text{A}}}\boldsymbol{\Lambda } {\mathbf {V}_{\text{A}}} }\right \|^{2}_{ {\mathcal {F}}} \\&\quad ~ \qquad \qquad \qquad \qquad \qquad\nonumber {{ {+ \, \beta _{\text{U}} \| {\mathbf {U}_{\text{A}}}\|^{2}_{ {\mathcal {F}}} + \beta _{\text{V}} \| {\mathbf {V}_{\text{A}}}\|^{2}_{ {\mathcal {F}}} ,}}}
\end{align}
where $\tilde{\mathbf{U}}_{\text{opt}}$ represents the estimate of $\mathbf{H}$, while $\tilde{\mathbf{V}}_{\text{opt}}$ corresponds to the data estimate $\widehat{\mathbf{X}}$. The quantity $\boldsymbol{\Lambda}$ refers to a diagonal matrix representing the large-scale fading factors associated with different users and it is given as $\boldsymbol{\Lambda} = \mathrm{Diag}\left(\left[\sqrt{\mathfrak{b}_1}, \sqrt{\mathfrak{b}_2}, \cdots, \sqrt{\mathfrak{b}_{K_U}}\right]\right)$, where $\mathfrak{b}_{K_U}$ accounts for the attenuation caused by path loss for the $K_U$-th user. The regularization parameters $\beta_{\text{U}}$ and $\beta_{\text{V}}$ are employed for integrating any prior information about $\mathbf{H}$ and $\mathbf{X}_b$, respectively. The selection of $\beta_{\text{U}}$ and $\beta_{\text{V}}$ is based on the condition $\beta_{\text{U}} = \beta_{\text{V}} = \sigma_v^{2}$, which stems from the fact that the entries of matrices $\mathbf{H}$ and $\mathbf{X}_b$ have an average power equal to unity. Note that the decomposition $\tilde{\mathbf{U}}_{\text{opt}} \boldsymbol{\Lambda} \tilde{\mathbf{V}}_{\text{opt}}$ lacks uniqueness; thus solving equation \eqref{lowrank-prob} does not guarantee a unique recovery of $\mathbf{X}_b$. Therefore, we define an invertible matrix $\mathbf{\Gamma} \in \mathbb{C}^{K_U \times K_U}$ for ensuring that $\mathbf{H}'\boldsymbol{\Lambda} = \tilde{\mathbf{U}}_{\text{opt}} \boldsymbol{\Lambda} \mathbf{\Gamma}$ and $\mathbf{X}_b = \mathbf{\Gamma}^{-1}\tilde{\mathbf{V}}_{\text{opt}}$, which means that a series of pilot signals is required for estimating $\mathbf{\Gamma}$ before proceeding with the estimation of $\mathbf{X}$. Therefore, after resolving equation \eqref{lowrank-prob}, we can obtain $\tilde{\mathbf{V}}_{\text{A}}^{\text{opt}} \triangleq[\tilde{\mathbf{V}}_{\text{A}_p}^{\text{opt}}, \tilde{\mathbf{V}}_{\text{A}_d}^{\text{opt}}] = \mathbf{\Gamma}[({\mathbf{X}_p^H}), ({\mathbf{X}_{d,b}^H})]$. Starting from the pilot phase, $\tilde{\mathbf{V}}_{\text{A}_p}^{\text{opt}} = \mathbf{\Gamma} ({\mathbf{X}_p^H})$, we can express $\mathbf{\Gamma}$ as
\begin{align}
    \mathbf{\Gamma} = \tilde{\mathbf{V}}_{\text{A}_p}^{\text{opt}}\mathbf{X}_p^H.
\end{align}
During the data phase, where $\tilde{\mathbf{V}}_{\text{A}_d}^{\text{opt}} = \mathbf{\Gamma} ({\mathbf{X}_{d,b}^H})$ and given that $\mathbf{\Gamma}$ is invertible, an estimate of $\mathbf{X}_{d,b}$ is given by
\begin{align}
    \widehat{\mathbf{X}}_{d,b} = \mathbf{\Gamma}^{-1} \tilde{\mathbf{V}}_{\text{A}_d}^{\text{opt}} = \mathbf{X}_p^H(\tilde{\mathbf{V}}_{\text{A}_p}^{\text{opt}})^{\dagger}\tilde{\mathbf{V}}_{\text{A}_d}^{\text{opt}}. \label{data_rals}
\end{align}
Thus, the RALS-SB algorithm iteratively invokes equation \eqref{lowrank-prob} for the factors $\mathbf{U}_{\text{A}}$ and $\mathbf{V}_{\text{A}}$, while keeping one parameter fixed at a time, as
\begin{align} \widehat { {\mathbf {V}}}_{\text{A}}=&\mathop {\mathrm {argmin}}_{ {\mathbf {V}}_{\text{A}}\in \mathbb {C}^{K_U\! \times \! \tau_{c}}} \left \|{ {\mathbf {Y}}_{\omega,b } - \widehat { {\mathbf {U}}}_{\text{A}} \boldsymbol{\Lambda } {\mathbf {V}}_{\text{A}} }\right \|^{2}_{ {\mathcal {F}}} + \beta_{\text{V}} \| {\mathbf {V}}_{\text{A}}\|^{2}_{ {\mathcal {F}}},\label{RALS_1}\\ \widehat { {\mathbf {U}}}_{\text{A}}=&\mathop {\mathrm {argmin}}_{ {\mathbf {U}}_{\text{A}}\in \mathbb {C}^{N_{RF} \! \times \! K_U}} \left \|{ {\mathbf {Y}}_{\omega,b } - {\mathbf {U}}_{\text{A}}\boldsymbol{\Lambda } \widehat { {\mathbf {V}}}_{\text{A}} }\right \|^{2}_{ {\mathcal {F}}} + \beta_{\text{U}} \| {\mathbf {U}}_{\text{A}}\|^{2}_{ {\mathcal {F}}}. \label{RALS_2}\end{align}
While the problem in equation \eqref{lowrank-prob} is non-convex, the sub-problems presented in equations \eqref{RALS_1} and \eqref{RALS_2} are convex in nature and can be efficiently addressed by regularized least squares methods. Thus, the final expression for $\widehat{\mathbf{H}}$ using the RALS-SB framework is given by \cite{jain2013low}
\begin{equation} \widehat { {\mathbf {H}}}_{\text{RALS-SB}}= {\tilde{\mathbf {U}}}^{\text{opt}}\boldsymbol{\Lambda } \tilde{\mathbf{V}}_{\text{A}_p}^{\text{opt}}\mathbf{X}_p^{H}. \label{RALS_SB}\end{equation}
It is worth noting that the MU-THz-ML channel estimator described in equation \eqref{ML-est} exclusively relies on the received signal associated with the pilots $\mathbf{X}_p$ only. Conversely, the MU-RALS-SB channel estimator proposed in equation \eqref{RALS_SB} leverages the received signal pertaining to both the pilots $\mathbf{X}_p$ and data $\mathbf{X}_d$. Furthermore, we may conclude from equations \eqref{data_rals} and \eqref{RALS_SB} that the inclusion of $\mathbf{X}_p$ serves the purpose of resolving the ambiguity associated with the non-unique matrix factorization solutions. To provide further insight, when $\mathbf{X}_p$ is fixed for a given $\tau_p$, the performance of the estimator can be further improved by increasing $N$. The initialization of RALS-SB is carried out as follows. The components of $\widehat{\mathbf{U}}_{\text{A}}$ and $\widehat{\mathbf{V}}_{\text{A}}$ are initially generated from i.i.d Gaussian random variables with zero-mean and unit-variance, followed by the substitution of pilot positions within $\widehat{\mathbf{V}}_{\text{A}}$ by the corresponding pilot signals. Thus, the RALS-SB algorithm iteratively switches between equations \eqref{expand_V} and \eqref{expand_U} until convergence, as outlined in Algorithm-1. A significant limitation of the RALS-SB framework lies in the selection of an appropriate regularization parameter, which can have a significant impact on the performance. Additionally, optimizing these parameters can be challenging. Moreover, finding the global maximum relies on appropriate initialization \cite{srivastava2021data}. Motivated by the efficiency of semi-blind methods, there is a clear incentive for us to explore this advanced approach. Therefore, the next section will introduce a novel MU-WD-SB scheme for estimating the CSI in MU THz massive MIMO systems.
\section{MU THz WD-SB Technique for CSI estimation in MU THz massive MIMO systems}\label{SB-derive}
\vspace{-3pt}
In this section, we introduce our novel MU WD-SB technique that leverages both the training symbols and the second-order statistical characteristics of the unknown data symbols. Note that the MU CSI  $\mathbf{H}$ for THz scenerio can be decomposed as
\begin{equation} \mathbf {H}=\mathbf {W}\mathbf {T}^{H}. \label{decomp} \end{equation}
This decomposition may be interpreted as $\text {SVD}(\mathbf {H})=\mathbf {S}\boldsymbol {\Sigma }\mathbf {V}^{H}$, and we can set $\mathbf {W}=\mathbf {S}\boldsymbol {\Sigma }$
as well as $\mathbf {T}=\mathbf {V}$. The estimation of the whitening matrix $\mathbf{W}\in \mathbb {C}^{N_{BS}\times K_{U}}$ relies on the statistical characteristics of the unknown data stream $\mathbf{X}_{d,b}$, where each element of the data beam has an average power of $P_d$. Additionally, we estimate the unitary matrix $\mathbf{T}\in \mathbb {C}^{K_{U}\times K_{U}}$ using the pilot symbols only. The pair of distinct phases of data transmission are visually illustrated in Fig. \ref{fig:frame-struc}. Since the unitary matrix $\mathbf{T}$ involves considerably fewer parameters than $\mathbf{H}$, the proposed semi-blind scheme exhibits notably enhanced estimation accuracy compared to the conventional training-based techniques. To start with, the blind estimation of the whitening matrix $\mathbf{W}$ is executed as follows. Since $\mathbf{T}$ is a unitary matrix, it satisfies the property $\mathbf {T}\mathbf {T}^{H}= \mathbf {T}^{H} \mathbf {T}=\mathbf {I}_{K_{U}}$. By leveraging these properties, it can be deduced from \eqref{decomp} that
\begin{equation}
    \mathbf {H} \mathbf {H}^{H}=\mathbf {W}\mathbf {W}^{H}. \label{H_W_relation}
\end{equation}
The output correlation matrix $\mathbf {R}_{Y} \in \mathbb {C}^{N_{BS} \times N_{BS}}$ from equation \eqref{data-model} may be defined as 
\begin{equation}
\mathbf {R}_{Y} \triangleq \mathbb {E} \big [\mathbf {W}_{\text {RF}}\mathbf {Y}_{d,b}(\mathbf {W}_{\text {RF}}\mathbf {Y}_{d,b})^{H}\big].\label{correlation}
\end{equation}
By adhering to equation \eqref{data-model} and making use of the properties mentioned in equation \eqref{H_W_relation}, the output covariance matrix can be further rearranged as
\begin{equation} \mathbf {R}_{Y}= N P_{d} \mathbf {H}\mathbf {H}^{H} + N\sigma ^{2} \mathbf {I}_{N_{BS}},\label{compute}\end{equation}
which can also be reformulated as
\begin{equation} \mathbf {W}\mathbf {W}^{H}= \mathbf {H}\mathbf {H}^{H}=\dfrac {\mathbf {R}_{Y}-N\sigma ^{2}\mathbf {I}_{N_{BS}}}{NP_{d}}.\label{rearrange}\end{equation}
Thus the output corresponding to the blind data symbols can be employed for estimating the output covariance $\mathbf {R}_{Y}$ as
$\widehat {\mathbf {R}}_{Y}= \mathbf {W}_{\text {RF}}\mathbf {Y}_{d,b}\mathbf {Y}_{d,b}^{H}\mathbf {W}^{H}_{\text {RF}}$. This estimation is resilient, since we have $\widehat{\mathbf{R}}_Y \to \mathbf{R}_Y$ with a high degree of probability, as the number of data vectors $N$ increases. Using the singular value decomposition (SVD) of \eqref{rearrange}, one can obtain
\begin{equation} \widehat {\mathbf{U}}\widehat {\boldsymbol{\Sigma }} \widehat {\mathbf{U}}^{H}=\text {SVD}\left ({\dfrac {\widehat {\mathbf {R}}_{Y}-N \sigma ^{2}\mathbf {I}_{N_{BS}}}{NP_{d}}}\right).\label{SVD1}\end{equation}
Upon exploiting the relationship described in equation \eqref{H_W_relation}, the blind estimation of the whitening matrix $\mathbf{W}$ can now be achieved by using the principal component analysis (PCA) approach \cite{jiang2011low}, yielding 
\begin{equation} \widehat {\mathbf {W}}=\widehat {\mathbf{U}}\widehat {\boldsymbol{\Sigma }}^{1/2}.\label{whit_esti}\end{equation}
\begin{algorithm}[t!]
\caption{MU RALS-SB framework}\label{algo1}
\textbf{Input:} ${\mathbf {Y}}_{\omega,b }$\\
\textbf{Initialize:} $\widehat {\mathbf {U}}_{\text{A}}$ and $\widehat {\mathbf {V}}_{\text{A}}$ \\
\textbf{repeat}\\
\phantom{beta} \textbf{for} $i=1:\tau_{c}$ \textbf{do}\\
\phantom{gamma}$\widehat {\mathbf {V}}_{\text{A}}(:,i) = \left(\boldsymbol{\Lambda }{\widehat{\mathbf{U}}}_{\text{A}}^{H} ({\text{Diag}}[\mathbf{W}_{\text{RF}}^H(:,i)]){\widehat{\mathbf{U}}}_{\text{A}}\boldsymbol{\Lambda } + \beta_{\text{V}}{\mathbf{I}}\right)^{-1}\,\\\ \phantom{beta}\phantom{alpha}\phantom{alpha}\phantom{alpha}\boldsymbol{\Lambda }{\widehat{\mathbf{U}}}_{\text{A}}^{H}({\text{Diag}}[{\mathbf{W}}^H_{\text{RF}}(:,i)]){\mathbf{Y}}_{\omega,b}(:,i)$\\
\phantom{beta}\textbf{end for}\\
\phantom{beta}\textbf{for} $n=1:N_{RF}$ \textbf{do}\\
\phantom{gamma}$\widehat {\mathbf {U}}_{\text{A}}(n,:) = {\mathbf{Y}}_{\omega,b}(n,:)\left({\text{Diag}}[{\mathbf{W}}^H_{\text{RF}}(n,:)]\right){\widehat{\mathbf{V}}}_{\text{A}}^{H} \boldsymbol{\Lambda }\,\\\phantom{alpha}\phantom{alpha}\phantom{beta} \left(\boldsymbol{\Lambda }{\widehat{\mathbf{V}}}_{\text{A}}({\text{Diag}}[{\mathbf{W}}^H_{\text{RF}}(n,:)]){\widehat{\mathbf{V}}}_{\text{A}}^{H}\boldsymbol{\Lambda }+ \beta_{\text{U}}{\mathbf{I}}\right)^{-1}$\\
\phantom{beta}\textbf{end for}\\
\textbf{until} Convergence
\end{algorithm}
Furthermore, as discussed above, the unitary matrix $\mathbf{T}$ is estimated by using pilot vectors only and therefore we can design it by solving the following constrained optimization problem:
\begin{align} &{\min _{\mathbf {T}}} \left \lVert{ \mathbf {W}_{\text {RF}}\mathbf {Y}_{p}-\widehat {\mathbf {W}}\mathbf {T}^{H}\mathbf {X}_{p}}\right \rVert ^{2}_{F} \notag \\&\text {s.t.}~\mathbf {T}\mathbf {T}^{H}=\mathbf {I}_{K_{U}}.\label{constraint}\end{align}
For an orthogonal pilot matrix $\mathbf {X}_{p}$ satisfying the relationship given in equation \eqref{pilot-orthog}, the closed-form solution of estimating the unitary matrix $\mathbf{T}$ for the aforementioned optimization problem is given by
\begin{equation} \widehat {\mathbf {T}}=\widehat {\mathbf {V}}_{\text{SB}}\widehat {\mathbf {U}}_{\text{SB}}^{H},\label{uni-est}\end{equation}
where the matrices $\widehat{\mathbf{V}}_{\text{SB}}$ and $\widehat{\mathbf{U}}_{\text{SB}}$ can be obtained by exploiting the SVD of the equation as
\begin{equation} \widehat {\mathbf {U}}_{\text{SB}}\widehat {\boldsymbol{\Sigma }}_{\text{SB}}\widehat {\mathbf {V}}_{\text{SB}}^{H}=\text {SVD}\left ({\widehat {\mathbf {W}}^{H} \mathbf {W}_{\text {RF}}\mathbf {Y}_{p}\mathbf{X}_p^H }\right).\label{SVD2}\end{equation}
Finally, the MU SB estimate of the channel matrix $\mathbf {H}$ obtained through our MU-WD-SB framework is given by
\begin{equation} \widehat {\mathbf {H}}_{\text {WD-SB}}=\widehat {\mathbf {W}} \widehat {\mathbf {T}}^{H}=\widehat {\mathbf{U}}\widehat {\boldsymbol{\Sigma }}^{1/2}\widehat {\mathbf {U}}_{\text{SB}}\widehat {\mathbf {V}}_{\text{SB}}^{H}.\label{final}\end{equation}
Algorithm \ref{algo2} summarizes the proposed MU semi-blind estimation technique employing our MU-WD-SB framework.
\subsection{Constrained CRLB for the MU-WD-SB scheme}\label{ccrlb}
In this section, we derive an expression for the constrained CRLB of the proposed MU-WD-SB framework under the assumption that there is no interference from other users. In the proposed semi-blind approach, the unitary matrix is determined through complex rotation constraints, as discussed in equation \eqref{constraint}. In light of this, the constrained CRLB paradigm is ideally suited for determining the relevant CRLB for such an estimator. It is also worth noting that this differs from the conventional CRLB derivation used for a linear Gaussian model in \cite{kay1993fundamentals}. To begin with, let us assume that the whitening matrix $\mathbf{W}$ is perfectly known, which is a reasonable assumption given that $\mathbf{W}$ can be estimated with a sufficiently high level of precision by transmitting a suitable number of spatially uncorrelated data symbols, as demonstrated by our simulation results. Additionally, the estimation of $\mathbf{W}$ does not contribute to the pilot overhead of the system, since the whitening matrix is estimated via blind estimation, as described in equation \eqref{whit_esti}. 

The parameter to be estimated is defined as $\boldsymbol{\xi}$. The formulation of $\boldsymbol{\xi}$ is derived from \cite{van1994cramer}, which emphasizes the necessity of considering the complex conjugate of each complex parameter.
Therefore, we can define $ \boldsymbol {\xi }\in \mathbb {C}^{2K_{U}^{2}\times 1}$ as
\begin{equation} \boldsymbol {\xi }= \begin{bmatrix} \text {vec}\big (\mathbf {T}\big)^{T} ~~\text {vec}\big (\mathbf {T}^{\ast }\big)^{T}\end{bmatrix}^{T}. \label{parameter}\end{equation}
This aspect may be interpreted as
\begin{equation} \boldsymbol {\xi } =\big [\mathbf {t}_{1}^{T},\mathbf {t}_{2}^{T},\ldots,\mathbf{t}_{{K_U}}^{T},\mathbf {t}_{1}^{H},\mathbf {t}_{2}^{H},\ldots,\mathbf{t}_{{K_U}}^{H} \big]^{T},\label{column-wise}\end{equation}
where $\mathbf {t}_i$ denotes the $i$th column of the unitary matrix $\mathbf{T}$.
\begin{algorithm}[t!]
\caption{MU-WD-SB framework}\label{algo2}
\textbf{Input:} Data Output $\mathbf {Y}_{d,b}$, pilot output $ \mathbf {Y}_{p}$, noise variance $\sigma ^{2}$, power $P_{d}$, pilot matrix $\mathbf{X}_p$\\
\textbf{Output:} $\widehat {\mathbf {H}}_{\text {WD-SB}}$
\begin{itemize}
    \item Obtain the estimate $\widehat {\mathbf {R}}_{Y}= \mathbf {W}_{\text {RF}}\mathbf {Y}_{d,b}\mathbf{Y}_{d,b}^H\mathbf {W}^{H}_{\text {RF}}$
    \item Evaluate $\widehat {\mathbf{U}}\widehat {\boldsymbol{\Sigma }} \widehat {\mathbf{U}}^{H}=\text {SVD}\left ({\dfrac {\widehat {\mathbf {R}}_{Y}-N \sigma ^{2}\mathbf {I}_{N_{BS}}}{NP_{d}}}\right)$
    \item Obtain the estimate $\widehat {\mathbf {W}}=\widehat {\mathbf{U}}\widehat {\boldsymbol{\Sigma }}^{1/2}$
    \item Evaluate $\widehat {\mathbf {U}}_{\text{SB}}\widehat {\boldsymbol{\Sigma }}_{\text{SB}}\widehat {\mathbf {V}}_{\text{SB}}^{H}=\text {SVD}\left ({\widehat {\mathbf {W}}^{H} \mathbf {W}_{\text {RF}}\mathbf {Y}_{p} \mathbf{X}_p^H}\right)$
    \item Obtain the estimate $\widehat {\mathbf {T}}=\widehat {\mathbf {V}}_{\text{SB}}\widehat {\mathbf {U}}_{\text{SB}}^{H},$
    \item Obtain the estimate $\widehat {\mathbf {H}}_{\text {WD-SB}}=\widehat {\mathbf {W}} \widehat {\mathbf {T}}^{H}$
    \item \textbf{return:} The estimate $\widehat {\mathbf {H}}_{\text {WD-SB}}$
\end{itemize}
\end{algorithm}
Now, using $\text {SVD}(\mathbf {H})=\mathbf {S}\boldsymbol {\Sigma }\mathbf {T}^{H}$, equation \eqref{pilot-rx} can be rewritten as
\begin{equation} \underbrace {\mathbf {S}^{H}\mathbf {W}_{\text {RF}}\mathbf {Y}_{p}}_{\tilde{\mathbf{Y}}}=\boldsymbol{\Sigma }\mathbf {T}^{H}\mathbf {X}_{p}+ \underbrace {\mathbf {S}^{H}\mathbf {W}_{\text {RF}}\mathbf {V}_{p}}_{\tilde{\mathbf{V}}},\end{equation}
and upon vectorizing, we obtain
\begin{equation} \text {vec}\big ({{\tilde{\mathbf{Y}}}^H} \big)=\Big [\boldsymbol{\Sigma }^*\otimes ({\mathbf{X}_p})^H\Big]\text {vec}\big (\mathbf {T}\big)+\text {vec}\big (\tilde{\mathbf{V}}^H \big).\label{vec-Yp}\end{equation}
To estimate the complex parameter vector  $\text {vec}\big (\mathbf {T}\big)$ as described in \eqref{parameter}, the unconstrained Fisher Information Matrix (FIM) can be found by exploiting \eqref{pilot-orthog}; a simplified version of this can be expressed as
\begin{align} \mathbf {C}=\dfrac {1}{\sigma ^{2}}\Big [\boldsymbol{\Sigma }^{H}\boldsymbol{\Sigma }\otimes \mathbf{X}_{p}\mathbf{X}_p^H\Big]=\dfrac {\tau_p P_{p}}{\sigma ^{2}}\Big [|\boldsymbol{\Sigma }|^{2}\otimes \mathbf {I}_{\tau_p}\Big], \label{unconst-FIM}\end{align}
where $\boldsymbol{\Sigma}^H \boldsymbol{\Sigma} = |\boldsymbol{\Sigma}|^2$. The FIM for the parameter vector $\text{vec}\big (\mathbf{T}^{\ast }\big)$, as defined in \eqref{parameter}, can be observed to be equivalent to \eqref{unconst-FIM}, given the assumption of pilot matrix orthogonality, as indicated in \eqref{pilot-orthog}. Consequently, the FIM $\mathbf{C}_{\xi }\in \mathbb{C}^{2{K_U}^2\times 2K_{U}^{2}}$ for $\boldsymbol{\xi}$ is defined as
\begin{equation} \mathbf {C}_{\xi }= \mathbf {I}_{2}\otimes \mathbf {C}=\dfrac {\tau_p P_{p}}{\sigma ^{2}}\Big [\mathbf {I}_{2}\otimes |\boldsymbol{\Sigma }|^{2}\otimes \mathbf {I}_{\tau_p}\Big].\label{FIM-cxi}\end{equation}
Since $\boldsymbol{\xi}$ is a parameter subject to specific constraints defined as
\begin{equation}   
\mathbf{t}_{i}^{H}\mathbf {t}_{i}= 1 \phantom{\beta}\phantom{\beta} 1 \leq i \leq K_U \label{constraint-1}\end{equation}
\begin{equation}
\mathbf {t}_{i}^{H}\mathbf {t}_{j}= 0 \phantom{\beta}\phantom{\beta} 1 \leq i< j \leq K_U, \label{constraint-2}
\end{equation}
we may construct a set of essential constraints for the constrained estimation problem as follows:
\vspace{-5pt}
\begin{align}\label{ess-cons}
\mathbf {f}\big (\boldsymbol {\xi }\big)=\big [\mathbf {t}_{1}^{H}\mathbf {t}_{1}-1,\mathbf {t}_{1}^{H}\mathbf {t}_{2},\mathbf {t}_{2}^{H}\mathbf {t}_{1},\mathbf {t}_{1}^{H}\mathbf {t}_{3},\ldots,\mathbf {t}_{K_{U}}^{H}\mathbf {t}_{K_{U}}-1\big]^{T}=\mathbf {0}.\end{align}
From \cite{stoica1998cramer}, it becomes plausible that the constrained CRLB for the parameter vector $\boldsymbol{\xi}$ is given by
\begin{equation} \mathbf{C}_{\text{T}}(\boldsymbol {\xi })=\mathbf {B}\big (\mathbf {B}^{H}\mathbf {C}_{\xi}\mathbf {B}\big)^{-1}\mathbf {B}^{H},\label{CCRLB_xi}\end{equation}
where $\mathbf {B}\in \mathbb {C}^{2K_{U}^{2}\times K_{U}^{2}}$ acts as an orthonormal basis for the null space of the matrix $\mathbf {J}\big (\boldsymbol {\xi}\big)$, which is formed by differentiating  $\mathbf {f}\big (\boldsymbol {\xi })$ and exploiting the principles of complex derivatives. As detailed in \cite{kay1993fundamentals}, $\mathbf {J}\big (\boldsymbol {\xi}\big)$ is given by
\begin{equation} \mathbf {J}\big (\boldsymbol {\xi}\big)\triangleq \left [{\frac {\partial \mathbf {f}\big (\boldsymbol {\xi }\big)}{\partial \boldsymbol {\xi }},~\frac {\partial \mathbf {f}\big (\boldsymbol {\xi }\big)}{\partial \boldsymbol {\xi }^{\ast }}}\right].\label{diff}\end{equation}
The procedure of obtaining the matrix $\mathbf{B}$ from $\mathbf {J}\big (\boldsymbol {\xi}\big)$ is provided in Appendix-\ref{append:A}. Furthermore, the matrix $\mathbf {J}\big (\boldsymbol {\xi}\big)\in \mathbb {C}^{K_{U}^{2}\times 2K_{U}^{2}}$ has rank $2K_{U}^{2} - K_{U}^{2}$, which can be justified by the fact that the size of $\boldsymbol{\xi}$ is $(2K_{U}^{2} \times 1)$ and the total number of constraints that are included in $\mathbf {f}\big (\boldsymbol {\xi }\big)$ is $K_{U}^{2}$. Let the matrix $\mathbf {B}$ be defined as $[\mathbf {B}_{1}^{T},\mathbf {B}_{2}^{T}]^{T}$ with $\mathbf {B}_{1},\mathbf {B}_{2}\in \mathbb {C}^{K^{2}_{U}\times K^{2}_{U}}$. Let the parameter vector for the MU massive MIMO THz channel matrix $\mathbf {H}$ be $\mathbf {h}=\text {vec}\big (\mathbf {H}^{T}\big)$ and $\mathbf{\Upsilon}=[\mathbf {W}\otimes \mathbf {I}_{K_{U}}]$ . Then, the vectorized parameter vector can be expressed as $\mathbf {h}= \mathbf{\Upsilon} \text {vec}\big (\mathbf {T}^{\ast }\big)$.
Hence, we can obtain the CRLB for the error covariance of the estimate of $\mathbf {h}$ as
\begin{align} \label{error-cov}
\mathbb {E}\left [{\big (\hat {\mathbf {h}}-\mathbf {h}\big)\big (\hat {\mathbf {h}}-\mathbf {h}\big)^{H}}\right]\geq&\dfrac {\sigma ^{2}}{P_{p}\tau_{p}}[\mathbf {S}\boldsymbol{\Sigma}\otimes \mathbf {I}_{K_{U}}]\mathbf {B}^{\ast }_{1}\tilde {\boldsymbol{\Sigma}}^{-1}\mathbf {B}_{1}^{T}\\ \nonumber &\times \,\big[\mathbf {S}\boldsymbol{\Sigma }\otimes \mathbf {I}_{K_{U}}]^{H} =\mathbf {C}_{\text{H}},\end{align}
where, the matrix $\tilde {\boldsymbol{\Sigma }}\in \mathbb {C}^{K^{2}_{U}\times K^{2}_{U}}$ is given by $\tilde {\boldsymbol{\Sigma }}=\text {diag}([2\sigma ^{2}_{1},\sigma ^{2}_{1}+\sigma ^{2}_{2},\sigma ^{2}_{2}+\sigma ^{2}_{1},2\sigma ^{2}_{2},\ldots,]).$
The procedure to derive \eqref{error-cov} is included in Appendix-\ref{append:B}. The element $\mathbf {C}_{\text{H}}[(\mathfrak{K}-1)K_{U}+l,(\mathfrak{K}-1)K_{U}+l]$ of the matrix $\mathbf {C}_{\text{H}} $ provides the MSE resulting from the semi-blind estimation of each element $\mathbf {H}(\mathfrak{K},l)$ of the MU massive MIMO THz channel matrix $\mathbf {H}$, which is further given by
\begin{align} \label{MSE_gen}
\mathbb{E}\big [\vert \widehat {\mathbf {H}}_{\text {WD-SB}}(\mathfrak{K},l)-\mathbf {H}(\mathfrak{K},l)\vert ^{2}\big]\geq&\dfrac {\sigma ^{2}}{\tau_p P_{p}}\sum _{\nu=1}^{K_U}\sum _{j=1}^{K_U}\dfrac {\sigma ^{2}_{\nu}}{\sigma ^{2}_{j}+\sigma ^{2}_{\nu}}\\ \nonumber &\times \,\big \vert \mathbf {S}(\mathfrak{K},\nu)\big \vert ^{2}\big \vert \mathbf {T}(l,j)\big \vert ^{2},\end{align}
where $\mathbf {S}(\mathfrak{K},\nu)$ and $\mathbf {T}(l,j)$ denote the $(\mathfrak{K},\nu)$th element of the matrix $\mathbf {S}$ and $(l,j)$th element of the matrix $\mathbf {T}$, respectively. Under the assumptions that the pilots are orthogonal \eqref{pilot-orthog} and the whitening matrix $\mathbf {W}$ is perfectly known, it can be readily seen that the weighing factor $\sigma ^{2}_{\nu}/(\sigma ^{2}_{j}+\sigma ^{2}_{\nu})$ of each term in \eqref{MSE_gen} leads to the net reduction of estimation error as compared to the conventional ML THz channel estimation and RALS-SB framework. Additionally, it can be noted that upon setting this weighing factor equal to unity. We arrive at the MSE bound derived for the conventional ML THz estimate in \eqref{ML-est}.
\subsection{MSE performance enhancement for the MU-WD-SB framework}\label{MSE-comp}
This subsection will present a comprehensive analysis of the performance improvement achieved by the MU-WD-SB estimator over the conventional ML estimator. For the proposed scheme, the parameter vector to be estimated, denoted by $\mathbf{h}$, is defined in accordance with \eqref{parameter} as
$\mathbf{h}=\left [{\text {vec}(\mathbf {H})^{T}, \text {vec}(\mathbf {H}^{\ast })^{T}}\right]^{T} \in \mathbb {C}^{2N_{BS}K_{U} \times 1}$. The unconstrained FIM for the parameter vector $\mathbf{h}$ can be calculated as $\mathbf {C}_h=\frac {\sigma ^{2}}{\tau_p P_{p}}\mathbf {I}_{2N_{BS}K_{U}}$. It is also worth noting that the total number of unconstrained real parameters in the MU THz massive MIMO channel matrix $\mathbf {H}$ is $2N_{BS}K_{U}$, resulting in an equivalent number of complex parameters as $N_{BS}K_{U}$. As discussed in equation \eqref{decomp}, the proposed MU-WD-SB framework splits the estimation process into a pair of distinct phases. The estimation of the whitening matrix $\mathbf{W}$ leverages the statistical properties of the data symbols only, which enables the MU-WD-SB framework to utilize the training matrix $\mathbf{X}_p$ solely for the estimation of the unitary matrix $\mathbf{T}$ consisting of only $K_U^2$ real parameters. This reduces the number of free parameters and enhances the accuracy of the channel estimation in the proposed scheme. As detailed in equations \eqref{parameter}-\eqref{ess-cons} and following a similar methodology, the bound of the estimation error $\boldsymbol {\epsilon }_h= \widehat{\mathbf{h}}-\mathbf{h}$ of the parameter $\mathbf{h}$ is determined as
\begin{equation} \mathbb {E}\big [\boldsymbol {\epsilon }_h \boldsymbol {\epsilon }_h^H \big]\geq \tilde {\mathbf {B}}\big (\tilde {\mathbf {B}}^{H}\mathbf {C}_h \tilde {\mathbf {B}}\big)^{-1}\tilde {\mathbf {B}}^{H},\label{error_bound}\end{equation}
where $\tilde{\mathbf {B}} \in \mathbb {C}^{2N_{BS}K_{U}\times K_{U}^{2}}$ denotes an orthonormal basis for the matrix $\mathbf {J}\big (\mathbf{h}\big)\in \mathbb {C}^{K_{U}^{2}\times 2N_{BS}K_{U}}$. Note that the size of the matrix  $\tilde{\mathbf {B}}$ can be justified by the fact that the rank of the matrix $\mathbf {J}\big ({\mathbf {h}}\big)$ is $2N_{BS}K_{U} - K_{U}^{2}$. Upon substituting $\mathbf {C}_h$ into \eqref{error_bound} and applying the trace operator to both sides of the result, we arrive
\begin{align} \mathbb {E}\left [{ \left \lVert{ \widehat {\mathbf {H}}_{\text {WD-SB}}-\mathbf {H}}\right \rVert ^{2}_{\mathcal{F}} }\right]\geq \frac {1}{2}\text {Tr}\left [{\big (\tilde {\mathbf {B}}^{H}\mathbf {C}_h \tilde {\mathbf {B}}\big)^{-1}}\right]=\dfrac {\sigma ^{2}}{2 \tau_p P_{p}}K_{U}^{2}.\end{align}
Here, it is evident that this bound is directly proportional to the number $K_{U}^2$ of free parameters in the unitary matrix $\mathbf {T}$. By contrast, the THz ML estimation technique computes an unconstrained estimate of the channel matrix $\mathbf {H}$. This method employs the training matrix ${\mathbf {X}_p}$ to estimate the entire channel matrix $\mathbf {H}$, which encompasses $2N_{BS} K_{U}$ real parameters. Consequently, the MSE bound for the conventional ML estimator is formulated as
\begin{equation} \mathbb {E}\bigg [\left \lVert{ \widehat {\mathbf {H}}_{\text {ML}}-\mathbf {H}}\right \rVert ^{2}_{\mathcal{F}} \bigg]=\frac {1}{2}\text {Tr}[\mathbf {C}_{ {h}}] = \dfrac {\sigma ^{2}K_{U}N_{BS}}{P_{p}\tau_p}.\label{ML-MSE}\end{equation}
When we compare the per parameter MSE bound of the proposed MU-WD-SB framework to that of MU-THz-ML, the proposed SB technique exhibits a bound of $\frac {\sigma ^{2} N K_{U}^{2}}{2P_{p} \tau_{p}}\left({\frac {1}{K_{U}N_{BS}}}\right)$. The bound reduces as the quantity $N_{BS}$ at the BS increases. By contrast, the bound for the ML technique is found to be $\frac {\sigma ^{2}K_{U}N_{BS}}{P_{p}\tau_{p}}\left({\frac {1}{K_{U}N_{BS}}}\right)$, which remains constant vs. $K_U$ and $N_{BS}$. Thus, the MSE improvement can be formulated as
\begin{align} \label{comp-MSE}
\mathcal {G}\leq \mathbb {E}\bigg [\left \lVert{ \widehat {\mathbf {H}}_{\text {ML}}-\mathbf {H}}\right \rVert ^{2}_{\mathcal{F}} \bigg]\bigg /\mathbb {E}\bigg [\left \lVert{ \widehat {\mathbf {H}}_{\text {WD-SB}}-\mathbf {H}}\right \rVert ^{2}_{\mathcal{F}} \bigg]=\dfrac {2N_{BS}}{K_{U}}.\end{align}
From equation \eqref{comp-MSE}, one can observe that the gain $\mathcal {G}$ increases with the number $N_{BS}$ of RAs at the BS. Additionally, it is apparent that by setting $N_{BS}=K_{U}$, the proposed MU-WD-SB estimation technique exceeds the performance of the conventional ML estimation technique by 3dB. This is because, the MU-WD-SB technique exploits the data symbols along with the pilot symbols to improve the CSI estimation performance. However, the ML technique only utilizes the pilot symbols which severely degrades the performance in THz regime.
\renewcommand{\arraystretch}{1.3}
\begin{table*}
    \centering
      \caption{MU-THz-ML Computational Complexity}

\begin{tabular}{|c|c|c|c|}
\hline
    \textbf{Operations} & Complex Multiplication $(X)$ & Complex Addition $(A)$ & Number of real floating point operations \\ \hline
    $(\widehat{\mathbf{X}}_p^H)^\dagger$ &  $K_U^3$ & - & $6X$ \\
    \hline
    $\mathbf{W}_{\text{RF}}\mathbf{Y}_p(\widehat{\mathbf{X}}_p^H)^\dagger$ &  $N_{BS}^2 K_U + N_{BS} K_U^2$ & $N_{BS}(N_{BS}-1)K_U + N_{BS}(K_U-1)K_U$ & $6X+2A$ \\
    \hline \hline
    $\widehat{\mathbf{H}}_{\text{ML}}$ &  $K_U^3 + N_{BS}^2 K_U + N_{BS} K_U^2$ & $N_{BS}(N_{BS}-1)K_U + N_{BS}(K_U-1)K_U$ & $6X+2A$ \\ \hline
    \end{tabular}
    
     \label{MLCom}
\end{table*}

\renewcommand{\arraystretch}{1.3}
\begin{table*} [hbt!]
    \centering
      \caption{MU-RALS-SB Computational Complexity}
  
\begin{tabular}{|c|c|c|c|c|}
\hline
    \textbf{Operation} & Repeat &Complex Multiplication $(X)$ & Complex Addition $(A)$ & Number of real floating point operations\\ \hline
    $\widehat{\mathbf{V}}_A$ & $i = 1 \cdots \tau_c$  &$K_U^3$ & $K_U^3 - K_U^2$ & $6X+2A+126K_U^3$ \\
    \hline
    $\widehat{\mathbf{U}}_A$ & $n=1 \cdots N_{\text{RF}}$ &$ K_U^3$ & $K_U^3 - K_U^2$ & $6X+2A + 126 K_U^3$ \\\hline \hline
     $\widehat{\mathbf{H}}_{\text{RALS-SB}}$ & - &$\tau_cK_U^3+N_{RF} K_U^3$ & $\tau_c (K_U^3 - K_U^2) + N_{RF}(K_U^3 - K_U^2)$&$6X+2A$ \\
    \hline
    \end{tabular}

     \label{RALS-SB} 
\end{table*}

\renewcommand{\arraystretch}{1.2}
\begin{table*}[hbt!]
    \centering
      \caption{MU-WD-SB Computational Complexity}
\scriptsize
\begin{tabular}{|c|c|c|c|}
\hline
    \textbf{Operation} & Complex Multiplications $(X)$ & Complex Additions $(A)$ & Number of real floating point operations\\ \hline
    $\widehat{\mathbf{W}}$ &  $N_{BS}^2 K_U N + K_U^3 + N_{BS}^2 K_U$ & $N_{BS}(N-1) K_U+ K_U^3 - K_U^2 + N_{BS}(N_{BS}-1)K_U$ & $6X + 2A + 126 N_{BS}^3$ for the SVD\\ \hline
    $\widehat{\mathbf{T}}$ & $N_{BS} K_U^2 + N_{BS}^3$ & $N_{BS}(K_U^2 - K_U) + (N_{BS}^3 - N_{BS}^2)$ & $6X + 2A + 126 K_U^3$ for the SVD \\ \hline \hline
    $\widehat{\mathbf{H}}_{\text{WD-SB}}$ &  \parbox{4.5cm}{$N_{BS}^2 K_U N + K_U^3 + N_{BS}^2 K_U+N_{BS} K_U^2 + N_{BS}^3$} & \parbox{7cm}{$N_{BS}(N-1) K_U+ K_U^3 - K_U^2 + N_{BS}(N_{BS}-1)K_U + N_{BS}(K_U^2 - K_U) + (N_{BS}^3 - N_{BS}^2)$} & $6X + 2A$\\ \hline
    \end{tabular}
  
     \label{WD-SB}
\end{table*}
\section{Computational Complexity}
This section details the computational complexity of the proposed MU-RALS-SB and MU-WD-SB schemes in comparison to the pilot-based MU-THz-ML scheme. The complexity is quantified in terms of real floating-point operations (flops), where a real flop corresponds to an addition, multiplication, or division. Tables \ref{MLCom}, \ref{RALS-SB}, and \ref{WD-SB} list the overall complexity of the MU-THz-ML, MU-RALS-SB, and MU-WD-SB estimates, along with various key computational steps. As described in \cite{kay1993statistical}, the MU-THz-ML scheme possesses a computational complexity of $\mathcal{O}(K_U^3 + N_{BS}^2 K_U + N_{BS} K_U^2)$ flops, primarily arising from the inversion and multiplication operations involved in the ML estimation process. The proposed MU-RALS-SB algorithm has a complexity of $\mathcal{O}(\tau_cK_U^3+N_{RF}K_U^3)$ flops, where the cubic dependence on $K_U$ stems from the iterative least squares optimization steps required for refining the channel estimate. Furthermore, the MU-WD-SB algorithm has a complexity of $\mathcal{O}(N_{BS}^2 K_U N + K_U^3 + N_{BS}^2 K_U+N_{BS} K_U^2 + N_{BS}^3)$ flops. This complexity arises due to the incorporation of singular value decomposition (SVD) for subspace processing, which enhances the estimation accuracy by leveraging the statistical characteristics of the received data. Among the three schemes, MU-THz-ML has the lowest computational complexity, but also the poorest performance. On the other hand, the MU-WD-SB technique exhibits cubic-order complexity due to the SVD operations, which exploit the statistical characteristics of the data symbols. However, this increased complexity translates into the highest performance in terms of SNR. The complexity ordering of the three schemes is given by $\mathcal{O}$ (MU-WD-SB) $>$ $\mathcal{O}$ (MU-RALS-SB) $>$ $\mathcal{O}$(MU-THz-ML) \cite{kay1993statistical}.
\section{Spatially sparse combining for MU THz massive MIMO systems}
In this subsection, our objective is to design THz hybrid MIMO RCs tailored for the MU scenario, aiming to maximize the spectral efficiency \cite{el2014spatially}, defined as
\begin{multline}
    \mathcal{R}(\mathbf{W}_{\text{RF}},\mathbf{W}_{\text{BB}}) = \text{log}_2 \left |\mathbf{I}_{K_U} + \frac{1}{\sigma^2_v K_U} \mathbf{R}_{\text{n}}^{-1} \mathbf{J}^H \mathbf{H} \mathbf{H}^H \mathbf{J} \right|, \label{specteffc}
\end{multline}
where the noise covariance matrix is given by $\mathbf{R}_{\text{n}} = \mathbf{J}^H \mathbf{J}$ and $\mathbf{J} = \mathbf{W}_{\text{RF}}\mathbf{W_{\text{BB}}}$. Previous studies, such as \cite{li2016robust} and \cite{zhu2017low}, have addressed the problem of separately designing the hybrid RCs of different users. Furthermore, they have designed the hybrid combiners under the assumption of perfect CSI knowledge at the receiver, which, in practical scenarios, is typically unavailable. Moreover, Morsali and Champagne \cite{morsali2019robust} designed hybrid RCs by considering norm-bounded channels to account for the characteristics of imperfect CSI. On the other hand, the proposed Bayesian learning-based design jointly optimizes the RCs across all users by leveraging the estimated channel knowledge from the proposed MU-WD-SB framework. Therefore, the problem of minimizing the mean squared error between the transmitted and received signals for the hybrid RCs $\mathbf{W}_{\text{RF}}$ and $\mathbf{W}_{\text{BB}}$ can be expressed as
\vspace{-5pt}
\begin{align}
    \big(\mathbf{W}_{\text{RF}}^{\text{opt}}, \mathbf{W}_{\text{BB}}^{\text{opt}}\big) = &\mathop{\text{arg min}} \limits_{{\mathbf{W}_{\text{RF}}, \mathbf{W}_{\text{BB}}}} \mathbb{E}\Big[\parallel\mathbf{y} - \mathbf{W}_{\text{BB}}^H\mathbf{W}_{\text{RF}}^H\mathbf{x}\parallel^2_2\Big],\notag  \\ & \text{s.t.} \quad |\mathbf{W}_{\text{RF}}(i,j)| = \frac{1}{\sqrt{N_{BS}}}.
\end{align}
Due to the non-convex constraint on the analog combiner, the direct optimization of the above cost-function is challenging. Thus, in order to streamline the problem, we initially discard the constraint imposed by the analog RC and examine a scenario of a fully digital RC. Consequently, the optimal MMSE solution, accounting for the fully digital configuration is given by
\begin{align}
    \mathbf{W}_{\text{MMSE}} = \widehat{\mathbf{H}}\left(\widehat{\mathbf{H}}^H\widehat{\mathbf{H}} + K_U \sigma_v^2 \mathbf{I}_{K_U}\right)^{-1}. \label{MMSE_eq} 
\end{align}
However, the MMSE RC obtained may not be directly factorizable into a product of RF and baseband RCs. Consequently, as suggested in \cite{el2014spatially}, the hybrid RC may be redefined as
\begin{align}
    \widetilde{\mathbf{W}}_{\text{BB}}^{\text{opt}} = &\mathop{\text{arg min}} \limits_{\widetilde{\mathbf{W}}_{\text{BB}}} \Big\Vert \mathbb{E}[\mathbf{y}\mathbf{y}^H]^{\frac{1}{2}} \left( \mathbf{W}_{\text{MMSE}} - \mathbf{G}_r \widetilde{\mathbf{W}}_{\text{BB}} \right) \Big\Vert_\mathcal{F}, \notag \\
    & \text{s.t.} \quad \parallel \text{diag}(\widetilde{\mathbf{W}}_{\text{BB}}\widetilde{\mathbf{W}}_{\text{BB}}^H) \parallel_0 = K_U, \label{optimal_combiner}
\end{align}
where the columns of the analog RC $\mathbf{W}_{\text{RF}}$ can be easily selected from the dictionary $\mathbf{G}_r$, which is defined as $\mathbf{G}_r = [\mathbf{a}_r{(\phi_1)}, \mathbf{a}_r{(\phi_2)}, \cdots, \mathbf{a}_r{(\phi_S)}] \in \mathbb{C}^{N_R \times S}$. The AoAs $\left\{\phi_s, \phantom{\beta} \forall \; 1\leq s \leq S\right\}$ span the angular range of $[0, \pi]$ and obey $\text{cos}(\phi_s) = \frac{2}{S}(s-1)-1$, where $S$ represents the angular grid size. Moreover, the constraint in equation \eqref{optimal_combiner}, arises from the fact that there can only be $K_U$ non-zero rows in the matrix $\widetilde{\mathbf{W}}_{\text{BB}}$, which leads to a simultaneous sparse structure. The next subsection delves into the procedure of Bayesian Learning based hybrid RC design for MU THz massive MIMO systems.
\subsection{Bayesian Learning-based Hybrid RC}
The Bayesian learning method involves assigning a parameterized Gaussian prior $p\left(\widetilde{\mathbf{W}}_{\text{BB}};\mathbf{\Omega}\right)$ to the baseband combiner matrix $\widetilde{\mathbf{W}}_{\text{BB}}$ as
\begin{align} p\left(\widetilde{\mathbf{W}}_{\text{BB}};\mathbf{\Omega}\right) & = \prod_{i=1}^S p\left(\widetilde{\mathbf{W}}_{\text{BB}}(i,:);\gamma_i\right) \notag \\ & = \prod_{i=1}^S \frac{1}{\pi \gamma_i} \text{exp}\left(-\frac{\parallel \widetilde{\mathbf{W}}_{\text{BB}}(i,:) \parallel^2}{\gamma_i}\right), \label{para_gau}
\end{align}
where $\mathbf{\Omega} = \text{diag}(\gamma_1, \gamma_2, \cdots, \gamma_S) \in \mathbb{R}^{S \times S}$ represents the diagonal matrix of hyperparameters. Consequently, the MMSE estimate of $\widetilde{\mathbf{W}}_{\text{BB}}$, denoted as $\boldsymbol{\mathcal{M}} \in \mathbb{C}^{S \times K_U}$, is accompanied by the corresponding error covariance matrix $\mathbf{\Pi} \in \mathbb{C}^{S \times S}$ and variance of approximation error $\sigma_a^2$ as 
\begin{align}
    \boldsymbol{\mathcal{M}} = \frac{1}{\sigma_a^2} \mathbf{\Pi} \mathbf{G}_r^H\mathbf{W}_{\text{MMSE}} \phantom{\beta} \text{and} \phantom{\beta} \mathbf{\Pi} = \left(\frac{1}{\sigma_a^2} \mathbf{G}_r^H \mathbf{G}_r + \mathbf{\Omega}^{-1}\right)^{-1}.
\end{align}
Moreover, it is evident that the MMSE estimate $\boldsymbol{\mathcal{M}}$ is influenced by hyperparameter matrix $\mathbf{\Omega}$. Notably, as $\gamma_i \to 0$, the corresponding row $\widetilde{\mathbf{W}}_{\text{BB}}(i,:) \to 0$. Consequently, the task of estimating $\widetilde{\mathbf{W}}_{\text{BB}}$ is analogous to estimating the associated hyperparameter vector $\boldsymbol{\gamma} = [\gamma_1, \cdots, \gamma_S]^T$. Thus, one can maximize the Bayesian evidence $p(\mathbf{W}_{\text{opt}};\mathbf{\Omega})$ by employing the iterative EM algorithm, which comprises two steps: the expectation step (E-step) and the maximization step (M-step). Moreover, the framework ensures likelihood maximization in each iteration and converges to a local optimum. To begin with, let $\widehat{\mathbf{\Omega}}^{(k-1)}$ represent the hyperparameter matrix estimate obtained from the $(k-1)$st iteration. During the E-step, we compute the average log-likelihood $\mathcal{L}\left(\mathbf{\Omega}|\widehat{\mathbf{\Omega}}^{(k-1)}\right)$ for the complete data set, which is expressed as
\begin{align}
\mathcal{L}\left(\mathbf{\Omega}|\widehat{\mathbf{\Omega}}^{(k-1)}\right) = \mathbb{E}_{\widetilde{\mathbf{W}}_{\text{BB}}|\mathbf{W}_{\text{opt}};\widehat{\mathbf{\Omega}}^{(k-1)}}\left\{\text{log} \phantom{\beta} p \left(\mathbf{W}_{\text{opt}},\widetilde{\mathbf{W}}_{\text{BB}};\mathbf{\Omega}\right)\right\}. \label{E-step_eq}
\end{align}
The M-step then proceeds to maximize the log-likelihood with respect to the hyperparameter vector $\boldsymbol{\gamma}$, resulting in
\begin{align}
    \hat{\boldsymbol{\gamma}}^{(k)} = \mathop{\text{arg max}}\limits_{\boldsymbol{\gamma}} \mathbb{E}\left\{\text{log} \phantom{\beta} p \left(\mathbf{W}_{\text{opt}}|\tilde{\mathbf{W}}_{\text{BB}}\right) + \text{log} \phantom{\beta} p \left(\widetilde{\mathbf{W}}_{\text{BB}};\mathbf{\Omega}\right) \right\}. \label{M-step_eqn}
\end{align}
From \eqref{M-step_eqn}, one can readily observe that the expression $p\left(\mathbf{W}_{\text{opt}}|\widetilde{\mathbf{W}}_{\text{BB}}\right)$ is independent of the hyperparameter $\boldsymbol{\gamma}$, which implies that it can be overlooked in the ensuing M-step. Consequently, the optimization problem can be reformulated
\begin{align}
    \hat{\boldsymbol{\gamma}}^{(k)} &= \mathop{\text{arg max}}\limits_{\boldsymbol{\gamma}} \mathbb{E}_{\tilde{\mathbf{W}}_{\text{BB}}|\mathbf{W}_{\text{opt}};\widehat{\mathbf{\Omega}}^{(k-1)}}\left\{\text{log} \phantom{\beta} p \left(\widetilde{\mathbf{W}}_{\text{BB}}; \mathbf{\Omega}\right)\right\} \notag \\ & \equiv \mathop{\text{arg max}}\limits_{\boldsymbol{\gamma}} \sum_{i=1}^S \bigg[-\log(\gamma_i) -\frac{\mathbb{E}(\parallel \widetilde{\mathbf{W}}_{\text{BB}}(i,:)\parallel_2^2)}{\gamma_i} \bigg] \notag \\ & \equiv \mathop{\text{arg max}}\limits_{\boldsymbol{\gamma}} \sum_{i=1}^S - \text{log}(\gamma_i) - \frac{\parallel \boldsymbol{\mathcal{M}}^{(k)}(i,:)\parallel^2+ K_U \mathbf{\Pi}_{(i,i)}^{(k)} }{\gamma_i},
\end{align}
which finally yields the update equation as 
\begin{align}
    \gamma_i^{(k)} = \frac{1}{K_U} \parallel \boldsymbol{\mathcal{M}}^{(k)}(:,i) \parallel^2 + \mathbf{\Pi}_{(i,i)}^{(k)}.
\end{align}
\begin{figure*}
	\centering
\subfloat[]{\includegraphics[scale=0.42]{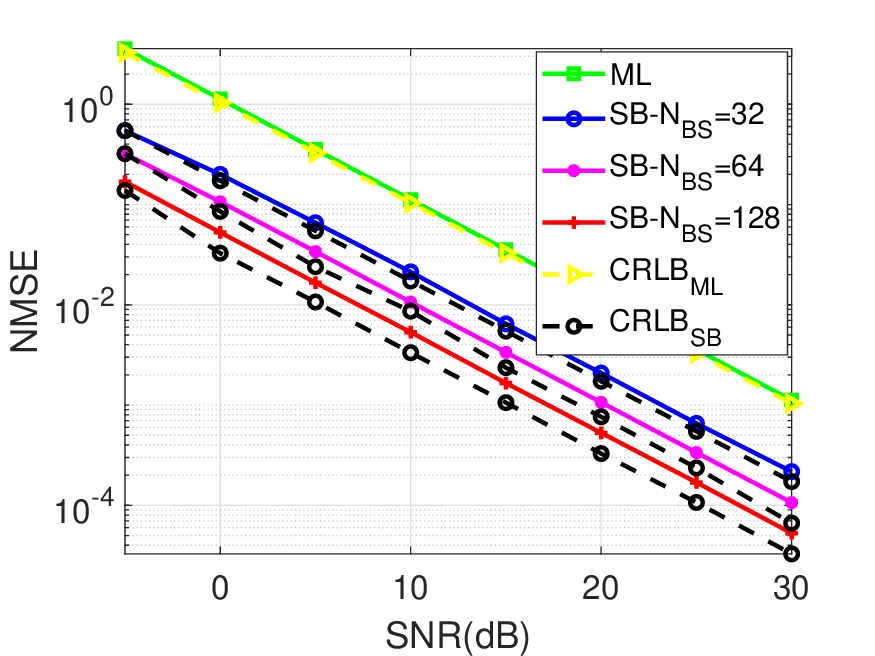}}
	\hfil
	\hspace{-10pt}\subfloat[]{\includegraphics[scale=0.42]{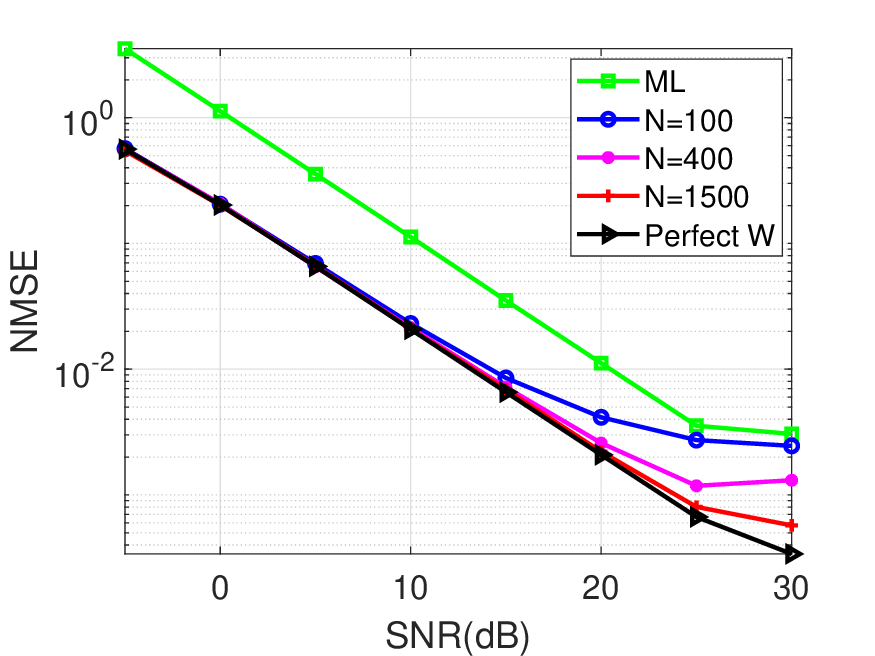}}
        \hfil
	\hspace{-10pt}\subfloat[]{\includegraphics[scale=0.43]{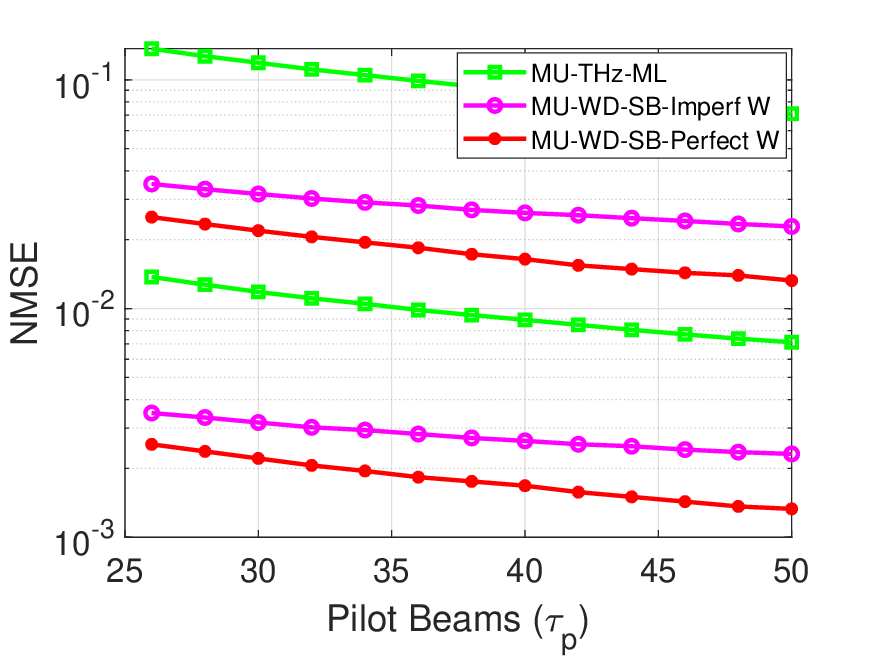}}
 \vspace{-5pt}
	\caption{NMSE versus SNR comparison between conventional MU-THz-ML estimation scheme and MU-WD-SB scheme $ \left(a\right) $ NMSE vs SNR for $N_{BS}\in \left\{32,64,128\right\}$ with $\mathbf{W}$ perfectly known along with the CRLB plots.  $ \left(b\right) $ NMSE vs SNR for $N\in \left\{100, 400, 1500\right\} $ with  $\mathbf{W}$ perfectly and imperfectly known $ \left(c\right) $ NMSE versus number of pilot beams for SNR$\in \left\{-5,5\right\}$dB.}\label{fig:SB-comp}
\end{figure*}
When the EM procedure converges, the optimal analog and digital RCs are given as
\begin{align}
    \mathbf{W}^H_{\text{RF}} = \mathbf{G}_r(:,S) \phantom{\beta} \text{and} \phantom{\beta} \mathbf{W}^H_{\text{BB}} = \widetilde{\mathbf{W}}_{\text{BB}}(S,:),
\end{align}
where $S$ denotes a set of indices for $N_{RF}$ hyperparameters having the largest value.
\section{Simulation Results}\label{simulation}
This section provides our performance analysis of the proposed MU-WD-SB technique, comparing it to other semi-blind techniques such as MU-RALS-SB and the conventional training-based technique. We use the NMSE as the evaluation metric for characterizing the efficacy of the MU-WD-SB technique, which is defined as $\left \Vert{ \widehat {\mathbf {H}}_{(.)}-\mathbf {H}}\right \Vert ^{2}_{\mathcal{F}}/\left \Vert{ \mathbf {H}}\right \Vert ^{2}_{\mathcal{F}}$. To compute the path gains corresponding to the LoS and NLoS components for the MU THz channel, we utilize equations \eqref{los_path} and \eqref{Nlos_path} respectively. The associated phase shifts are generated using a uniform probability distribution spanning the range $(-\pi,\pi]$. The RA array maintains a constant inter-antenna spacing of $d_{r} = \frac {\lambda }{2}$. The molecular absorption coefficient appearing in equations \eqref{los_path}, \eqref{Nlos_path} is calculated using the HITRAN database \cite{rothman2009hitran}. The transmission distance $d$ is configured to be $15$m, while the central frequency $f$ of the THz system is set to $0.3$ THz \cite{dovelos2021channel} with bandwidth $B$ = $5$GHz. We consider an indoor office scenario with a molecular composition of $20.9$\% oxygen, $78.1$\% nitrogen and $1$\% water vapor, where the temperature and pressure are set to $298$ K and $1$ atm, respectively. The MU THz channel consists of $N_{cl} = 4$ clusters with one LoS component and $3$ NLoS components \cite{tarboush2024cross}, where the NLoS components possess first-order reflection, which is given by \eqref{ray_rough}. We consider $3$ scatterers that can be encountered in a general office scenario, which are detailed in Table \ref{materials}.
\begin{table}[hbt!] 
\centering
\caption{List of materials used for the simulation environment \cite{piesiewicz2007scattering}}
\begin{tabular}{|l|c|c|c|}
\hline
    \textbf{Material Type} & $\sigma$(in mm) & $\varsigma$ (in cm$^{-1}$) & $n$ \\\hline
    Plaster s1 &  $0.05$ & $10$ &  $2$ \\ \hline
    Gypsum plaster &  $0.13$ & $38$ & $1.4$ \\  \hline
    Plaster s2 &  $0.15$ & $10$ & $2$ \\  \hline
    \end{tabular}
     \label{materials}
\end{table}
For simplicity, we have considered only one diffused ray with an AoA of $\phi_r$, but the results can be extended to multiple diffused rays as well. We set the combined TA and RA gain $B_r$ to $26$ dBi and generate the MU THz massive MIMO channel using equations \eqref{channel-vec}-\eqref{NLOS-channel}, while considering the above mentioned parameters. Moreover, all the results are generated for the single BS MU-massive MIMO THz uplink. Additionally, the data symbols are randomly generated from a QPSK constellation with an average power of unity. The SNR value in decibels (dB) is calculated as $\text{SNR(dB)}= 10 \log _{10}\displaystyle \left(\frac {1}{\sigma^{2}}\right)$, assuming unity average power. Fig. \ref{fig:SB-comp}(a) illustrates the performance enhancement achieved by the proposed MU-WD-SB technique compared to the training-based MU-THz-ML scheme. It is evident from the figure that the performance of the proposed MU-WD-SB framework closely approximates the C-CRLB, as detailed in Section-\ref{ccrlb}. As a result, when considering different values of $N_{BS}$ as $N_{BS} \in \left\{32,64, 128\right\}$,  while keeping $K_U$ fixed at $12$ and $N_{RF} = 16$ \cite{hossain2021thz}, we observe NMSE performance improvements of $7.3$ dB, $10.28$ dB, and $13.29$ dB, respectively. Theoretically, the MU-WD-SB scheme exhibits an NMSE performance improvement of $10\log_{10}\left(\frac{2N_{BS}}{K_{U}}\right)$ compared to the MU-THz-SB scheme, as derived in Section IV. This improvement is visually evident in the figure, validating the effectiveness of the proposed scheme. Fig. \ref{fig:SB-comp}(b) illustrates the NMSE vs SNR performance for the proposed MU-WD-SB scheme under two scenarios: when the whitening matrix $\mathbf{W}$ is perfectly known and when it is imperfectly known. The parameters considered for the simulation include $N_{BS} = 64$, $N_{RF} = 16$ with $K_U=12$. Notably, as the number of data symbols $N$ increases, the estimation performance of the proposed MU-WD-SB improves significantly, approaching the performance achieved when $\mathbf{W}$ is perfectly known. This observation aligns with equations \eqref{rearrange} and \eqref{whit_esti}, where the estimated correlation matrix $\widehat{\mathbf{R}}_{Y}$ gradually approaches the true correlation matrix $\mathbf{R}_{Y}$. Upon increasing the number of data symbols $N$. This, in turn, results in the estimated whitening matrix $\widehat{\mathbf{W}}$ approaching the true whitening matrix $\mathbf{W}$, which confirms the efficacy of the proposed scheme. Note that, at high SNR, the proposed scheme encounters a flooring effect, which arises due to error-propagation in $\mathbf{R}_y$. These propagated errors degrade the precision in estimation of the left singular vectors of $\mathbf{U}$, a critical step in the construction of the whitening matrix $\mathbf{W}$ (as stated in Eq. \eqref{whit_esti}). However, this flooring effect diminishes as the sample size $N$ increases. With larger $N$, the statistical estimates of $\mathbf{R}_y$ improve, leading to superior performance across the entire SNR range. Fig. \ref{fig:SB-comp}(c) shows the NMSE performance versus the number of pilot beams $\tau_p$ for both the MU-THz-SB and MU-WD-SB schemes, considering constant SNR conditions and both perfect as well as imperfect whitening matrix $\mathbf{W}$ scenarios. The simulation parameters are set as $N_{BS} = 64$, $N_{RF} = 16$ and the number of data streams $N$ is fixed at $1000$.
\begin{figure*}
	\centering
\subfloat[]{\includegraphics[scale=0.42]{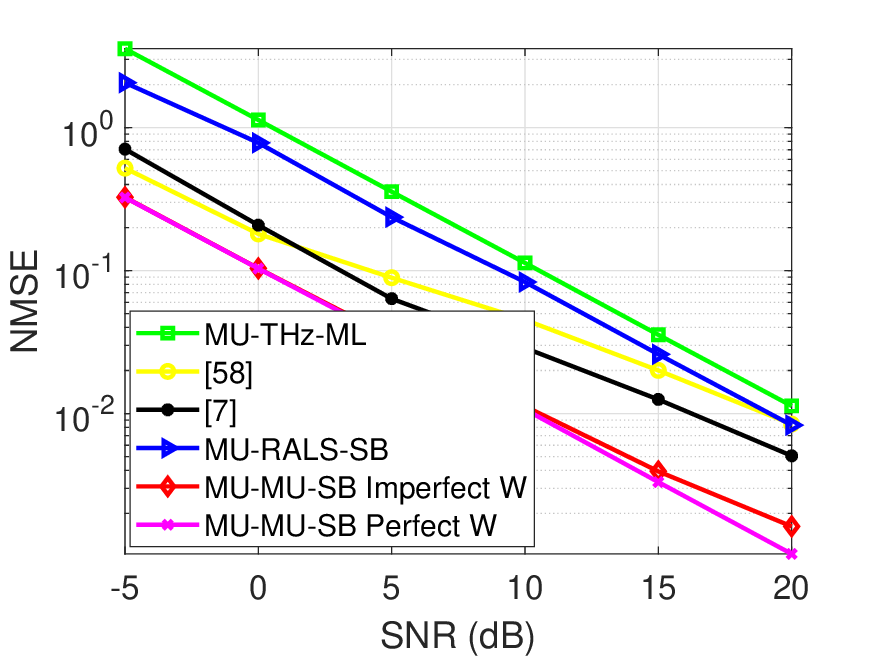}}
	\hfil
	\hspace{-10pt}\subfloat[]{\includegraphics[scale=0.42]{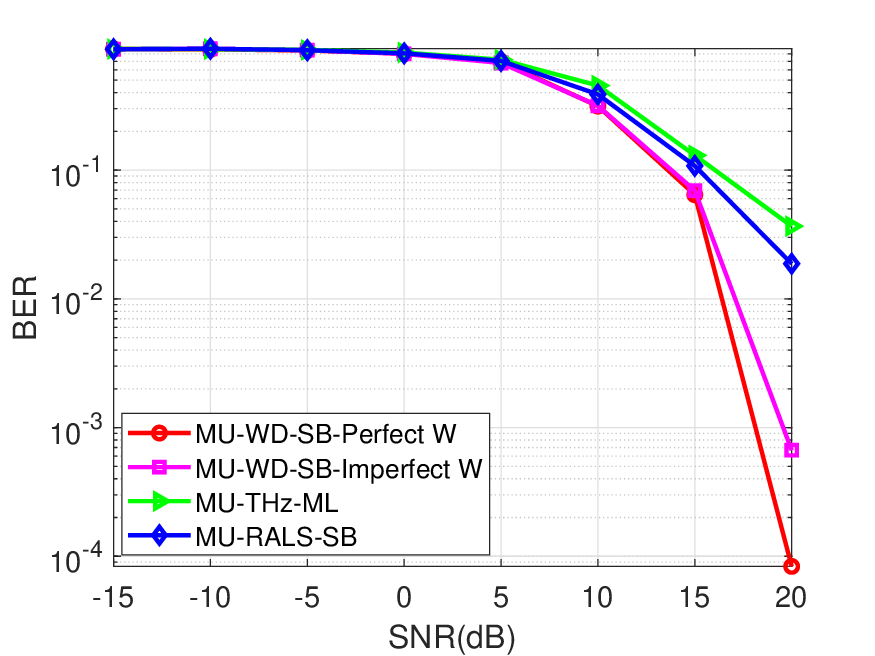}}
 	\hspace{-10pt}\subfloat[]{\includegraphics[scale=0.42]{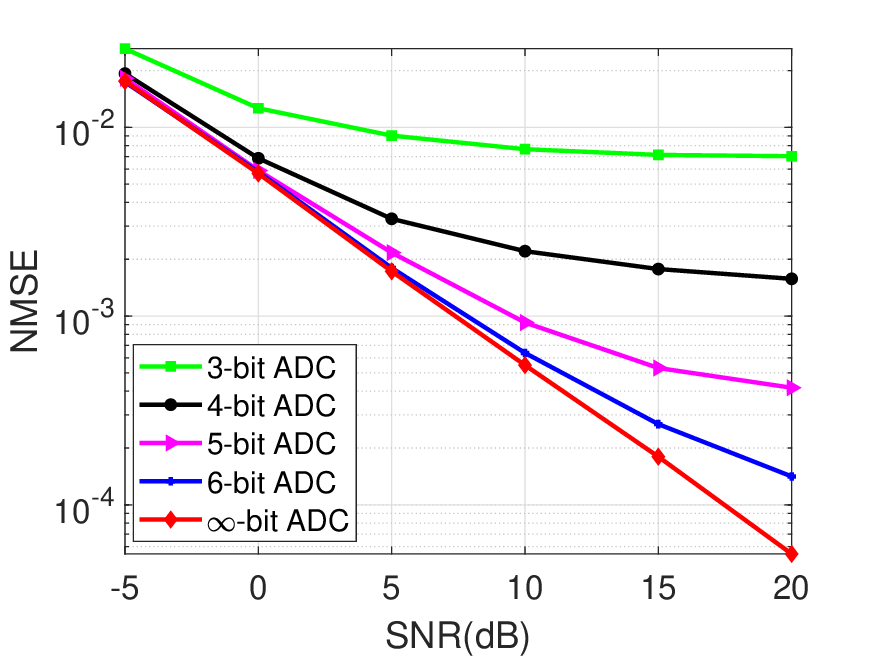}}
        \hfil
        \vspace{-5pt}
	\caption{Comparison plots between different channel estimation schemes $ \left(a\right) $ NMSE versus SNR comparison between MU-THz-ML, MU-RALS-SB and the proposed MU-WD-SB scheme. $ \left(b\right) $ BER vs SNR plot compared with ML estimation scheme along with  $\mathbf{W}$ perfectly and imperfectly known. $ \left(c\right) $ NMSE vs SNR comparison for the MU THz massive MIMO system considering low-resolution ADCs.}
	\label{fig:Diff_comparison-plots}
\end{figure*}
The results show that the estimation performance of both schemes improves upon increasing the number of pilot beams, as expected. Furthermore, it is evident that the proposed MU-WD-SB scheme requires fewer pilots than the conventional MU-THz-ML scheme for attaining a specific NMSE performance level. This reduction in the number of pilots required for attaining a given NMSE performance is a significant advantage of the proposed MU-WD-SB scheme, reaffirming its effectiveness. 
\subsection{Performance comparison with other estimation techniques}
In this subsection, we conduct a comprehensive performance analysis of the proposed MU-WD-SB scheme and contrast it to that of the conventional training based MU-RALS-SB, as well as sparse estimation techniques such as OMP \cite{tropp2007signal} and SBL \cite{srivastava2022hybrid}. Fig. \ref{fig:Diff_comparison-plots}(a) illustrates the NMSE \textit{vs.} SNR comparison for different channel estimation schemes, revealing that the proposed MU-WD-SB scheme outperforms the aforementioned design alternatives. The simulation parameters include $N_{BS} = 64$, $N_{RF} = 16$, $N = 1000$, $K_U = 12$ \cite{tarboush2024cross}. Notably, the proposed MU-WD-SB estimation scheme achieves the desired NMSE performance at significantly lower SNR levels than the conventional estimation schemes. Furthermore, it outperforms the MU-RALS-SB technique, which faces challenges due to large matrix inversions. Thus, as the THz channel's dimension becomes large due to a large number of antennas, the practicality of the MU-RALS-SB scheme erodes. Additionally, a notable advantage of the proposed MU-WD-SB scheme is its superior performance in comparison to OMP. This is due to the fact that while the OMP is known to be sensitive to threshold settings, the SBL framework requires complex priors, which can reduce their overall effectiveness. The proposed MU-WD-SB scheme intelligently overcomes these limitations, which renders it extremely well suited for practical implementations. The effectiveness of the proposed MU-WD-SB scheme is demonstrated through a BER \textit{vs.} SNR plot. This is shown in Fig. \ref{fig:Diff_comparison-plots}(b) in comparison it to different channel estimation schemes. Notably, the proposed MU-WD-SB scheme outperforms both the conventional MU-THz-ML and semi-blind MU-RALS-SB schemes, underscoring its strong data detection capability.
\subsection{Effects of low-Resolution ADCs and empirical cumulative distribution function} 
\begin{figure*}
	\centering
\subfloat[]{\includegraphics[scale=0.42]{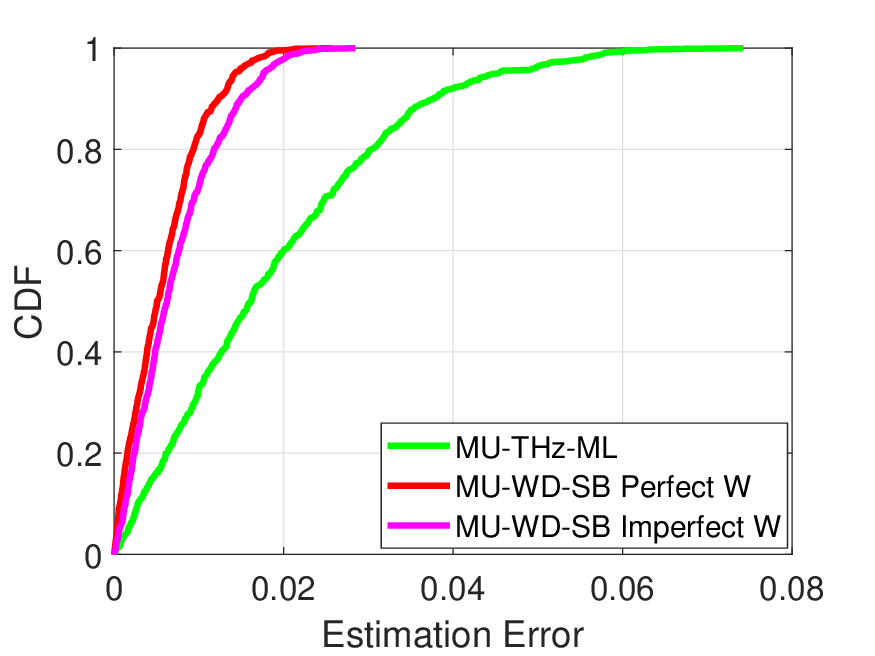}}
	\hfil
	\hspace{-10pt}\subfloat[]{\includegraphics[scale=0.42]{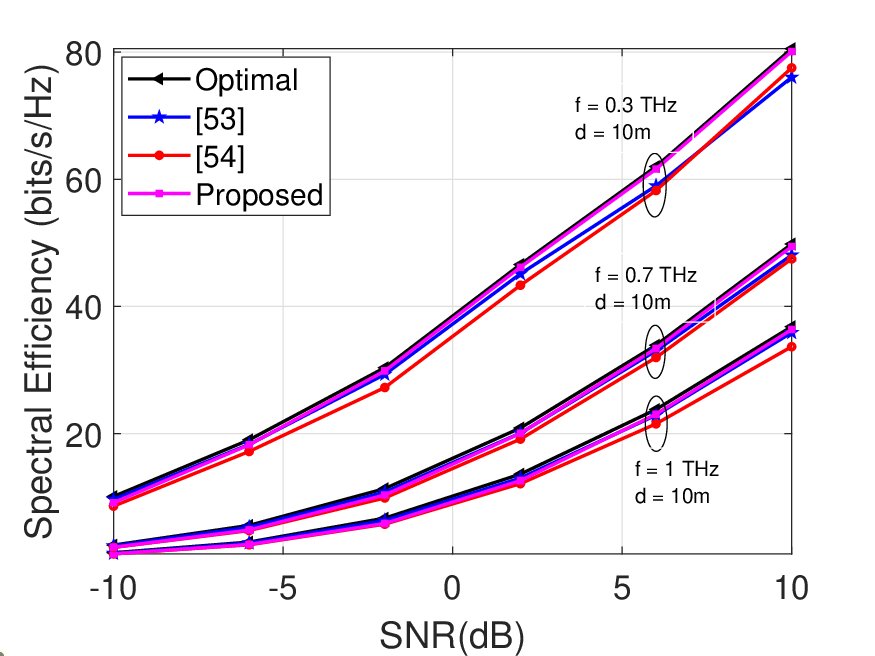}}
 	\hfil
 \hspace{-10pt}\subfloat[]{\includegraphics[scale=0.42]{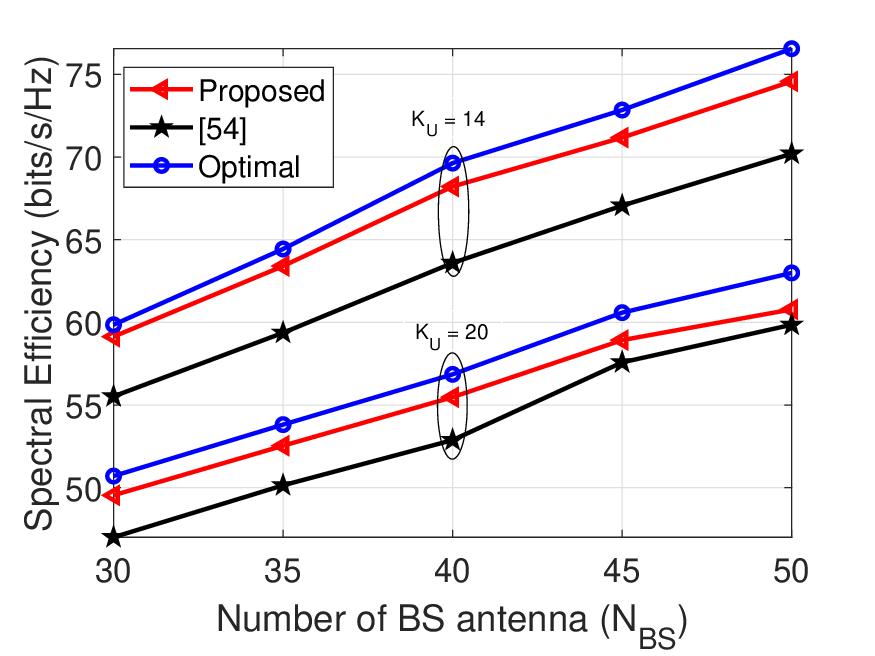}}
 \vspace{-5pt}
	\caption{$ \left(a\right) $ Empirical CDF for channel estimation error of MU-THz-ML and proposed MU-WD-SB with perfect and imperfect $\mathbf{W}$ $ \left(b\right) $ Spectral Efficiency vs SNR comparison for the MU massive MIMO THz system with different frequencies and distances $\left(c\right)$ Spectral Efficiency vs number of BS $N_{BS}$ antennas with different pilots.}
	\label{spectralefficiency}
\end{figure*}

Fig. \ref{fig:Diff_comparison-plots}(c) illustrates the impact of employing low-resolution ADCs on the estimated CSI performance for the proposed MU-WD-SB scheme. The quantized pilot output, derived from equation \eqref{pilot-rx} is denoted as $\mathbf{Y}_{p}^{\text{q}} = \mathcal{Q}(\mathbf{Y}_p)$, where $\mathbf{Y}_{p}^{\text{q}}(i) = \mathcal{Q}(\text{Real}[\mathbf{Y}_p(i)]) + j\mathcal{Q}(\text{Imag}[\mathbf{Y}_p(i)])$ \cite{ding2018bayesian}. The operator $\mathcal{Q}(.)$ represents the element-wise quantization for a uniform step size, and the number of quantization levels can be represented as $N_{\text{q}} = 2^b$. Moreover, the quantized pilot output after incorporating the quantized noise is given by
\begin{align}
    \mathbf{Y}_{p}^{\text{q}} = \mathbf{W}^H_{\text{RF}}\mathbf{H}\mathbf{X}_p^H + \mathbf{W}^H_{\text{RF}}\mathbf{V}_p + \mathbf{W}^H_{\text{RF}} \mathbf{V}_{\text{q}}.
\end{align}
One can readily observe from the figure that when 6-bit ADC resolution is considered, the NMSE of the proposed MU-WD-SB scheme is nearly identical to that achieved by $\infty$-bit resolution. However, within the low-SNR range of $-5$dB to $5$dB, the NMSE obtained through $4$-bit and $5$-bit ADC resolution remains acceptable. This observation underscores the practicality of the proposed MU-WD-SB scheme, in the face of practical low-resolution ADCs having a low power consumption. Fig. \ref{spectralefficiency}(a) compares the CDFs of estimation errors for training-based and semi-blind estimation schemes, namely for the MU-THz-ML and for the proposed MU-WD-SB schemes under the scenarios of both perfectly and imperfectly known whitening matrix. %Estimation errors are indicative of scheme effectiveness, and CDFs help visualize their capabilities in improving estimation performance.%
The estimation error is defined as $\boldsymbol{\epsilon}_h = \hat{\mathbf{h}}-\mathbf{h}$ for the parameter vector $\mathbf{h} = \text{vec}(\mathbf{H})$. The CDF \cite{katagiri2022radio} of $\boldsymbol{\epsilon }_h= (\boldsymbol{\epsilon }_{{h}_1},\boldsymbol {\epsilon }_{{h}_2},\cdots,\boldsymbol {\epsilon }_{{h}_{N_{BS} K_{U}}})$ is calculated
\begin{equation}
\hat {F}_{k}(\boldsymbol{\epsilon }_{{{h}}_r}) = \frac {1}{N_{BS} K_{U}} \sum _{\jmath=0}^{N_{BS} K_{U}} P(\hat{h}_\jmath-h_\jmath),\label{ECDF}
\end{equation}
where $P(\hat{h}_\jmath-h_\jmath)$ is the indicator function of
\begin{align} 
P(\hat{h}_\jmath-h_\jmath) = \begin{cases} 1 & \text {if}\;(\boldsymbol {\epsilon }_{{h}_\jmath} \leq \boldsymbol {\epsilon }_{{{h}}_r}) \\ 0 & \text {otherwise}, \end{cases}\label{indicator}
\end{align}
where $\boldsymbol {\epsilon }_{{h}_\jmath} = (\hat{h}_\jmath-h_\jmath) $ and $\boldsymbol {\epsilon }_{{{h}}_r}$ is the threshold error value. The figure distinctly illustrates that the estimation errors associated with the MU-THz-ML scheme are significantly higher when compared to the proposed MU-WD-SB scheme. This observation underscores the effectiveness of the proposed MU-WD-SB scheme in delivering high estimation accuracy judicious.
\subsection{Hybrid combiner design}
Fig. \ref{spectralefficiency}(b) illustrates the SE as a function of SNR for the parameters of $N_{BS} = 64$, $N_{RF} = 16$, $K = 14$, $B = 26 \, \text{dB}$. Notably, our proposed hybrid RC relying on realistic estimated CSI closely approaches the optimal fully digital MMSE beamformer, which serves as the benchmark for our simulations. This observation underscores the effectiveness of our proposed hybrid RC, as well as of our proposed MU-WD-SB framework. Moreover, we have compared the performance of our proposed RC to that of other approaches \cite{li2016robust, zhu2017low}, which demonstrates a significant performance improvement and proves the efficacy of the proposed RC. To extend our analysis further, we plot the SE at three different frequencies of $f \in \left\{0.3, 0.7, 1\right\}$ THz. The hybrid combiner designs in \cite{li2016robust} \cite{zhu2017low} often assume fixed beam selection strategies, which limit adaptability in dynamic channel conditions, especially in the presence of low-resolution ADCs. By contrast, our SBL-based hybrid combiner jointly optimizes the RF and baseband processing stages by iteratively learning the optimal weight vectors, leading to improved beam alignment, reduced quantization noise effects, and enhanced spectral efficiency. Additionally, SBL provides a more probabilistic framework for beam selection, allowing the combiner to dynamically adapt to varying user positions and propagation conditions, which is crucial for MU-massive MIMO uplink systems. As depicted in figure \ref{spectralefficiency}(b), the SE at $1$ THz recorded for a fixed distance of $d = 10$ m is lower, which is attributed to the increased free-space loss, as described in equation \eqref{fre-space}. Fig. \ref{spectralefficiency}(c) presents another noteworthy result by considering the SE as a function of $N_{BS}$. As the number of BS RAs increases, there is a consistent improvement in the SE. It underscores the importance of considering the RA configurations and the allocation of resources for optimizing the system performance attained. Moreover, the results highlight the efficacy and adaptability of our proposed combiner in comparison to other approaches \cite{zhu2017low}. Recent studies have explored the application of DMAs in beamforming for downlink multi-user systems \cite{chen2024energy} and uplink massive MIMO systems \cite{shlezinger2019dynamic}. Zhang et al., \cite{zhang2025tensor} propose a tensor-based channel estimation scheme for hybrid MIMO-OFDM systems with DMAs, exploiting channel sparsity to reduce training overhead without iterative refinement. DMAs eliminate the need for additional analog circuitry by integrating analog processing directly into their metamaterial elements. In hybrid beamforming, phase shifters are typically used to interconnect a large number of antenna elements with fewer RF chains, enabling a combination of analog and digital processing. However, these phase shifters significantly increase the power consumption and hardware overhead. While both DMA-based and hybrid beamforming share the goal of reducing the number of RF chains compared to fully digital architectures, DMAs provide a more efficient and streamlined solution by eliminating the need for phase-shifting networks, thereby achieving higher energy efficiency and reduced hardware complexity. Potential future work can include the incorporation of DMAs into advanced beamforming architectures to further enhance the system performance attained.
\section{Conclusions}\label{conclusion}
A novel semi-blind framework was conceived for MU THz CSI acquisition tailored for a single-cell scenario. The various THz frequency band characteristics such as molecular absorption loss, reflection losses and path gains are also taken into account. We capitalized on the second-order statistics of the unknown data symbols, which effectively reduced the amount of training data required for channel estimation, while enhancing the accuracy. A significant advantage of our proposed scheme is its reduced training overhead. Aditionally, to assess the estimation performance, we derived the C-CRLB, which also validated the efficacy of our proposed scheme. Additionally, we provided analytical insights for characterizing the improvements in terms of the NMSE and BER performances attained. Moreover, our simulation results offer empirical evidence of the enhanced performance achieved by the proposed MU-WD-SB scheme in comparison to both training-based and semi-blind schemes in the THz regime.
\begin{appendices}
\section{Calculation of  Matrix $\mathbf{B}$}
\label{append:A}
The matrix $\mathbf{J}\big(\boldsymbol{\xi}\big)$ can be constructed by differentiating the expression in \eqref{ess-cons}, leveraging the properties of complex derivatives as discussed in \cite{kay1993fundamentals}, which is seen below
\begin{align} \mathbf {J}\big (\boldsymbol {\xi }\big)=&\begin{bmatrix} \mathbf {t}_{1}^{H}&\mathbf {0}& \mathbf {0}& \cdots & \mathbf {t}_{1}^{T}& \mathbf {0}& \mathbf {0} & \cdots \\ \mathbf {0}& \mathbf {t}_{1}^{H}&\mathbf {0}&\cdots & \mathbf {t}_{2}^{T}&\mathbf {0}&\mathbf {0}&\cdots \\ \mathbf {t}_{2}^{H}&\mathbf {0}&\mathbf {0}&\cdots &\mathbf {0}&\mathbf {t}_{1}^{T}&\mathbf {0}&\cdots \\ \mathbf {0}&\mathbf {t}_{2}^{H}&\mathbf {0}& \cdots &\mathbf {0}& \mathbf {t}_{2}^{T} &\mathbf {0}&\cdots \\ \mathbf {t}_{3}^{H}&\mathbf {0}&\mathbf {0}&\cdots &\mathbf {0}&\mathbf {0}&\mathbf {t}_{1}^{T} &\cdots \\ \mathbf {0}&\mathbf {0}&\mathbf {t}_{1}^{H}&\cdots &\mathbf {t}_{3}^{T}&\mathbf {0}&\mathbf {0}&\cdots \\ \vdots &\vdots &\vdots &\ddots &\vdots & \vdots & \vdots &\ddots \end{bmatrix}\quad. \end{align}
In accordance with the principles outlined in \cite{van1994cramer} for the C-CRLB, it is crucial to consider both the complex parameter and its complex conjugate counterpart for each complex parameter. When examining the set of non-redundant constraints in \eqref{ess-cons}, it is evident that the total number of constraints equals to $K_U^2 + 2 \binom{K_U}{2} = K_U^2$. This automatically confirms that the number of free parameters is $2K_U^2 - K_U^2 = K_U^2$. Consequently, by exploiting the properties of the matrix $\mathbf{B} \in \mathbb{C}^{2K_U^2 \times K_U^2}$ as an orthonormal basis for the null-space of the matrix $\mathbf{J}(\boldsymbol{\xi}) \in \mathbb{C}^{K_U^2 \times 2K_U^2}$, the matrix $\mathbf{B}$ can be formulated as
\begin{align}
\mathbf {B}=&\dfrac {1}{\sqrt {2}} \begin{bmatrix} \mathbf {t}_{1}&\mathbf {0}& \mathbf {t}_{2}&\mathbf {0}& \mathbf {t}_{3}&\mathbf {0}&\cdots \\ \mathbf {0}& \mathbf {t}_{1}&\mathbf {0}& \mathbf {t}_{2}&\mathbf {0}&\mathbf {0}&\cdots \\ \mathbf {0}&\mathbf {0}&\mathbf {0}&\mathbf {0}&\mathbf {0}&\mathbf {t}_{1}&\cdots \\ \vdots &\vdots & \vdots &\vdots &\vdots &\vdots &\ddots \\ -\mathbf {t}^{\ast }_{1}&-\mathbf {t}^{\ast }_{2}& \mathbf {0}&\mathbf {0}& \mathbf {0}&\mathbf {t}^{\ast }_{3}&\cdots \\ \mathbf {0}&\mathbf {0}&-\mathbf {t}^{\ast }_{1}&\mathbf {t}^{\ast }_{2}&\mathbf {0}&\mathbf {0}&\cdots \\ \mathbf {0}&\mathbf {0}&\mathbf {0}&\mathbf {0}&-\mathbf {t}^{\ast }_{1}&\mathbf {0}&\cdots \\ \vdots &\vdots & \vdots &\vdots &\vdots &\vdots &\ddots \end{bmatrix}.\end{align}
\section{Constrained CRLB for the proposed MU-WD-SB technique}
\label{append:B}
By substituting \eqref{FIM-cxi} into \eqref{CCRLB_xi}, one can simplify the matrix $\mathbf{B}^{H}\mathbf{C}_{\xi}\mathbf{B}$ as
\begin{equation} \mathbf {B}^{H}\mathbf {C}_{\xi }\mathbf {B} = \dfrac {P_{p}\tau_{p}}{\sigma ^{2}}\tilde {\boldsymbol{\Sigma }}.\label{simplify}\end{equation}
As discussed in Section IV-A, the matrix $\mathbf {B}$ can be partitioned as $\mathbf {B}=[\mathbf {B}_{1}^{T},\mathbf {B}_{2}^{T}]^{T}$, with $\mathbf{B}_1$, $\mathbf{B}_2 \in \mathbb{C}^{K_U^2 \times K_U^2}$.
Thus, the matrix $\mathbf {C}_{\text{T}}(\boldsymbol {\xi })$
can be partitioned into a block matrix with each block being of size $K_{U}^{2}\times K_{U}^{2}$ as
\begin{align} \label{partitioning}
\mathbf {C}_{\text{T}}(\boldsymbol {\xi })= \begin{bmatrix} \left [\mathbf {C}_{\text{T}}(\boldsymbol {\xi })\right]_{1,1} &\quad \left [\mathbf {C}_{\text{T}}(\boldsymbol {\xi })\right]_{1,2} \\ \left [\mathbf {C}_{\text{T}}(\boldsymbol {\xi })\right]_{2,1} &\quad \left [\mathbf {C}_{\text{T}}(\boldsymbol {\xi })\right]_{2,2} \end{bmatrix}\!.\end{align}
From \eqref{FIM-cxi} and \eqref{CCRLB_xi}, the C-CRLB for $\text {vec}\big (\mathbf {T}\big)$  is derived upon exploiting \eqref{partitioning} and the partitioned $\mathbf{B}$ matrix as
\begin{align}
\mathbf {C}_{\text{T}}(\boldsymbol {\xi })_{1,1}=(\sigma ^{2}/(P_{p}\tau_{p}))\mathbf {B}_{1}\tilde {\boldsymbol{\Sigma }}^{-1}\mathbf {B}_{1}^{H}.
\end{align}
Furthermore, as discussed, the vectorized parameter vector is given as $\mathbf{h} = \mathbf{\Upsilon}\text{vec}(\mathbf{T}^*)$. The error covariance of the MU THz channel $\mathbf{h}$ is bounded as
\begin{align} \label{mse-derived}
&\hspace {-1pc}\mathbb {E}\left [{\big (\hat {\mathbf {h}}-\mathbf {h}\big)\big (\hat {\mathbf {h}}-\mathbf {h}\big)^{H}}\right]\notag\\\geq &\nonumber[\mathbf {W}\otimes \mathbf {I}_{K_{U}}]\big ([\mathbf {C}_{\text{T}}(\boldsymbol {\xi })]_{1,1}\big)^{\ast }[\mathbf {W}\otimes \mathbf {I}_{K_{U}}]^{H}\\ =&\dfrac {\sigma ^{2}}{P_{p}\tau_{p}}[\mathbf {S}\boldsymbol{\Sigma }\otimes \mathbf {I}_{K_{U}}]\mathbf {B}^{\ast }_{1}\tilde {\boldsymbol{\Sigma }}^{-1}\mathbf {B}_{1}^{T}[\mathbf {S}\boldsymbol{\Sigma }\otimes \mathbf {I}_{K_{U}}]^{H} =\mathbf {C}_{\text {H}}.\end{align}
\end{appendices}
\vspace{-1.8\baselineskip}
\bibliographystyle{IEEEtran}
\bibliography{References}
\end{document}